\documentclass[journal]{IEEEtran}
\usepackage[scr=boondox]{mathalfa}
\usepackage{pdflscape}
\usepackage[nocomma]{optidef} % For nice optimization models: https://mirror.math.princeton.edu/pub/CTAN/macros/latex/contrib/optidef/optidef.pdf
\usepackage{lipsum}
\usepackage[dvipsnames]{xcolor}
\usepackage[normalem]{ulem}
\usepackage[utf8]{inputenc}
\usepackage{multirow, multicol, makecell, booktabs, graphicx}
\usepackage{afterpage}
\usepackage{placeins}     % For \FloatBarrier
\usepackage{dsfont}
\usepackage{adjustbox}
\usepackage{url}
\usepackage[utf8]{inputenc}
\usepackage{amsmath,bm}
\usepackage{hyperref} %Hyperreferences
\hypersetup{
     colorlinks   = true,
     citecolor    = blue,
     linkcolor = blue,
     urlcolor = blue
}
\usepackage[switch]{lineno}
\usepackage{bbm}
\usepackage{amsthm}
\usepackage{enumerate}
\usepackage{float}
\usepackage{stfloats}
\usepackage[numbers]{natbib}
\setcitestyle{numbers,square}
\usepackage[english]{babel}
\usepackage{cases}
\usepackage{array}
\usepackage{amsthm}
\usepackage{amsmath}
\usepackage{amssymb}

\theoremstyle{definition}

\theoremstyle{definition}

\theoremstyle{definition}

\theoremstyle{definition}

\theoremstyle{definition}

\usepackage{braket}
\usepackage{textgreek}
\usepackage{mathtools, cuted}
\usepackage{amsfonts}
\usepackage{algorithm}
\usepackage[noend]{algpseudocode}
\usepackage{hhline}
\usepackage{float}
\usepackage[normalem]{ulem} % For strikthrough text
\usepackage{booktabs}
\usepackage{tabularx}

\algnewcommand\algorithmicinput{\textbf{Input:}}
\algnewcommand\Input{\item[\algorithmicinput]}
\algnewcommand\algorithmicoutput{\textbf{Output:}}
\algnewcommand\Output{\item[\algorithmicoutput]}

\usepackage [english]{babel}
\usepackage [autostyle, english = american]{csquotes}
\MakeOuterQuote{"}
\usepackage{hhline}
\usepackage{enumitem}
\usepackage{float}

\usepackage{caption}

\usepackage{subcaption}
\usepackage{amssymb}
\usepackage{siunitx}
\usepackage{tabstackengine}
\newcommand\AddLabel[1]{%
  \refstepcounter{equation}% increment equation counter
  (\theequation)% print equation number
  \label{#1}% give the equation a \label
}
\newcolumntype{C}{>{\hfil$\displaystyle}c<{$\hfil}}
\newcolumntype{R}{>{\hfil$\displaystyle}r<{$}}
\newcolumntype{L}{>{$\displaystyle}l<{$\hfil}}

\newcolumntype{Z}{>{\hfill$\displaystyle}X<{$\hfill}}
\newcolumntype{E}{>{\hfill$\displaystyle}X<{$}}
\newcolumntype{K}{>{$\displaystyle}X<{$\hfill}}

\usepackage{mathtools}
\DeclarePairedDelimiter\ceil{\lceil}{\rceil}

\newfloat{model}{tbp}{lom}
\floatname{model}{Model}

\newcommand{\integers}{\mathbb{Z}}

\newcommand{\Jobs}{\mathcal{J}}

\newcommand{\Machines}{\mathcal{M}}

\newcommand{\Graph}{\mathcal{G}}
\newcommand{\Nodes}{\mathcal{N}}
\newcommand{\Arcs}{\mathcal{A}}
\newcommand{\InitialSatets}{\Lambda}
\newcommand{\FinalStates}{\Omega}
\newcommand{\inistate}{\sigma}
\newcommand{\finalstate}{\varphi}
\newcommand{\makespan}{C_{\max}}
\newcommand{\Horizon}{\mathcal{H}}

\newcommand{\jm}{{jm}}
\newcommand{\jb}{{jb}}

\renewcommand{\bm}{{bm}}

\newcommand{\Batches}{\mathcal{B}}

\newcommand{\Families}{\mathcal{F}}

\newcommand{\machineStateFamily}{{\mathfrak{f}}}

\newcommand{\cumulLoad}{{\mathscr{l}}}

\newcommand{\ptime}{\delta}  % Processing time
\newcommand{\weight}{\omega}  % Priority
\newcommand{\releaseTime}{r}  % Release time
\newcommand{\size}{n}  % Size

\newcommand{\pp}[1]{\left(#1\right)}

\newcommand{\StartOf}{\texttt{startOf}}
\newcommand{\EndOf}{\texttt{endOf}}
\newcommand{\PresenceOf}{\texttt{presenceOf}}

\newcommand{\StartBeforeStart}{\texttt{startBeforeStart}}
\newcommand{\NoOverlap}{\texttt{noOverlap}}
\newcommand{\Synchronize}{\texttt{synchronize}}

\newcommand{\Alternative}{\texttt{alternative}}
\newcommand{\StateFunction}{\texttt{state}}

\newcommand{\Permutation}{\text{Perm}}
\newcommand{\pulse}{\text{pulse}}
\newcommand{\AlwaysEqual}{\texttt{alwaysEqual}}

\newcommand{\Automaton}{\texttt{automaton}}

\newcommand{\Element}{\texttt{element}}
\newcommand{\ExactlyOne}{\texttt{exactlyOne}}

\newcommand{\nt}[1]{{#1}}
\newcommand{\rt}[1]{{#1}}

\begin{document}

\title{Parallel Batch Scheduling With Incompatible Job Families Via Constraint Programming}

\author{
        Jorge A. Huertas,
        Pascal Van Hentenryck
    \thanks{
            This research was partly supported by the NSF \href{https://www.ai4opt.org}{\textit{AI Institute for Advances in Optimization}} (Award 2112533) \textit{(Corresponding author: Jorge A. Huertas.)}
    }% 
    \thanks{
            J.A. Huertas, and P. Van Hentenryck are with the \href{https://www.gatech.edu}{\textit{Georgia Institute of Technology}}, Atlanta, GA, USA (e-mails: \href{mailto:huertas.ja@gatech.edu}{huertas.ja@gatech.edu}, \href{mailto:pvh@gatech.edu}{pvh@gatech.edu}).
    }
}

%\markboth{IEEE TRANSACTIONS ON SEMICONDUCTOR MANUFACTURING}
%{Huertas and Van Hentenryck: Incompatible P-batch Scheduling Via CP}


\maketitle

% \linenumbers
\begin{abstract}
This paper addresses the incompatible case of parallel batch scheduling, where compatible jobs belong to the same family, and jobs from different families cannot be processed together in the same batch. \nt{The state-of-the-art} constraint programming (CP) model for this problem \nt{relies on specific functions and global constraints only available in a well established commercial CP solver}. This paper \nt{expands the literature around this problem by} propos\nt{ing four} new CP models \nt{that can be implemented in commercial and open-source solvers: a new model that relies on automaton constraints, and three alternative models that integrate assignment and scheduling decisions with different strategies and global constraints. Extensive computational experiments on standard test cases under multiple objectives and multiple solvers demonstrate the implementation flexibility and competitive performance of the proposed models.}
\end{abstract}


%%%%%%%%%%%%%%%%%%%%%%%%%%%%%%%%%%%%%%%%%%%%%%%%%%%%%%%%%%%%%%%%%%%%%%%%
% Introduction
%%%%%%%%%%%%%%%%%%%%%%%%%%%%%%%%%%%%%%%%%%%%%%%%%%%%%%%%%%%%%%%%%%%%%%%%

\section{Introduction} \label{sec: introduction}

In the current and highly competitive landscape of the manufacturing industry, companies are under growing pressure to minimize production costs and reduce cycle times. A crucial approach to achieving these goals is performing certain manufacturing processes in batches \cite{Monch2011-Survey, Fowler2022-SurveyP-Batching}. In \textit{parallel batching} (p-batch), jobs inside a batch are processed in parallel at the same time, allowing to reduce their cycle times \cite{Fowler2022-SurveyP-Batching, Monch2013-FabBook, Uzsoy1995}. 

P-batch is a common problem that appears in many contexts such as metal processing \cite{Liu2022}, truck delivery \cite{Cakici2013}, automobile gear manufacturing \cite{Gokhale2011}, additive manufacturing (AM) \cite{Nascimento2023, Nascimento2024, Cakici2024}, composite manufacturing \cite{Tang2020}, milk processing \cite{Escobet2019}, and many more \cite{Fowler2022-SurveyP-Batching, PottsKovalyov2000-BatchSchedulingReview}. Nonetheless, it has been mostly studied in the semiconductor manufacturing (SM) process, which is the most complex manufacturing process in the world and consists of two stages \cite{Monch2011-Survey, Fowler2022-SurveyP-Batching, Monch2013-FabBook}. The first stage is the \textit{front-end}, where the integrated circuits (ICs) are built on the silicon wafers in a wafer fabrication facility (wafer fab). The second stage is the \textit{back-end}, where the individual ICs are assembled (A) and tested (T) before final packaging in an AT fab. \cite{Fowler2022-SurveyP-Batching, Monch2013-FabBook}.

In the back-end, p-batching has been typically studied in the burn-in operations of the testing phase, where the individual ICs are subject to thermal stress over an extended period to detect latent defects \cite{Fowler2022-SurveyP-Batching, Kim2011}. The test of an individual IC is a job that requires a minimum processing time, predefined ahead of time. Due to the prolonged processing times, jobs are batched to be put in the oven, and the processing time of the batch is the longest of the jobs in it. In this setting, jobs can remain in the oven beyond their required time but cannot be removed before the batch is completed \cite{Uzsoy1994, Trindade2018}. This scheduling problem belongs to the \textit{compatible} case of p-batch, where all the jobs are compatible and can therefore be batched together \cite{Fowler2022-SurveyP-Batching}.

A more complex p-batch problem arises in the front-end of the SM process, in the diffusion area of the wafer fab, where dopants are introduced into the silicon wafers at high temperatures to alter their electrical properties. This is achieved using diffusion furnaces that operate in batches, applying constant heat on the wafers for long periods, typically 10-12 hours \cite{Kopp2020-FabTestbed}. These prolonged processing times are essential to ensure the thermal stability required for an even distribution of dopants on the wafers \cite{Ham2017-CP-p-batch-incompatible}. Due to the chemical nature of the process, not all the jobs can be batched together. Instead, \textit{incompatible job families} are defined by the distinct "recipes" of the jobs, which dictate unique dopants and thermal conditions \cite{Fowler2022-SurveyP-Batching}. Jobs with the same recipe are said to be compatible and therefore grouped into the same family; jobs in the same family also have the same processing time \cite{Monch2011-Survey, Wu2022-DiffusionFurnacesScheduling}. Jobs of different families are said to be \textit{incompatible} and cannot be processed together in the same batch, as this may result in unwanted cross-contamination in the chemical process. Hence, the scheduling problem in the diffusion area belongs to the incompatible case of p-batch, where the batches can only contain jobs of the same family \cite{Fowler2022-SurveyP-Batching, Ham2017-CP-p-batch-incompatible}. Furthermore, due to the risk of burning the wafers or damaging the chemical process, jobs cannot remain in the furnace beyond their processing time. Moreover, the processing of a batch cannot be interrupted to include or remove jobs mid-process, as this would disrupt the thermal stability and jeopardize the uniformity of the diffusion \cite{Ham2017-CP-p-batch-incompatible}. Thus, jobs inside a batch must start and end together.

Existing exact methods in the literature for p-batching include mixed-integer programming (MIP) \cite{Cakici2013, Ham2017-CP-p-batch-incompatible} and Constraint Programming (CP) models \cite{Nascimento2023, Ham2017-CP-p-batch-incompatible}. MIP models do not scale well and are outperformed by the CP approaches \cite{Fowler2022-SurveyP-Batching, Ham2017-CP-p-batch-incompatible}. However, the \nt{existing} CP model \nt{for p-batching with incompatible jobs is solver-dependent as it relies on specific functions and global constraints only available in a well know\rt{n} commercial CP solver.} This paper \nt{expands the body of knowledge around this problem} by proposing \nt{four} new CP models \nt{that can be implemented in other solvers: one model that relies on automaton constraints, available in different open-source solvers; and other three models integrate assignment and scheduling decisions with different strategies and global constraints, available in many commercial and open-source solvers.}
Furthermore, the paper conducts thorough computational experiments that demonstrate the \nt{implementation flexibility of the proposed models, and compares them under different objectives.}

The remainder of this paper is organized as follows. Section \ref{sec: lit review} presents a literature review on p-batching. Section \ref{sec: contributions} clearly outlines our contributions. Section \ref{sec: problem description} formally presents the description of the problem we address. Section \ref{sec: existing models} presents \nt{the existing solver-dependent CP model for p-batching}. Section \ref{sec: proposed models} presents the \nt{four} proposed CP models. Section \ref{sec: experiments and results} presents our computational experiments and their results. Finally, Section \ref{sec: conclusion} presents the conclusions and outlines future lines of research.

%%%%%%%%%%%%%%%%%%%%%%%%%%%%%%%%%%%%%%%%%%%%%%%%%%%%%%%%%%%%%%%%%%%%%%%%
% Literature review
%%%%%%%%%%%%%%%%%%%%%%%%%%%%%%%%%%%%%%%%%%%%%%%%%%%%%%%%%%%%%%%%%%%%%%%%

\section{Literature Review} \label{sec: lit review}

\rt{Bottlenecks in semiconductor manufacturing are dynamic and can occur in various areas depending on the process technology, product mix, and equipment availability \cite{Monch2013-FabBook}. Common areas that become bottlenecks in a wafer fab are the photolithography \cite{Wang2011_Photolithography}, ion implantation \cite{huertas2025s-batch}, and diffusion \cite{Ham2017-CP-p-batch-incompatible, Wu2022-DiffusionFurnacesScheduling} areas. This paper focuses on the diffusion area where the bottlenecks appear due the batching nature of the process to overcome the long processing times.}

The scheduling literature distinguishes two types of p-batch: \textit{compatible} and \textit{incompatible} p-batch \cite{Fowler2022-SurveyP-Batching, Monch2011-Survey, Monch2013-FabBook}. In the compatible case, all the jobs are compatible and can therefore be batched together. In contrast, in the incompatible case, jobs are grouped into \textit{families}, compatible jobs belong to the same family, and jobs of different families cannot be processed together in the same batch \cite{Monch2013-FabBook,Fowler2022-SurveyP-Batching, Uzsoy1995}.

The standard classification scheme for scheduling problems provided by \citet{Graham1979} is $\alpha|\beta|\gamma$, where $\alpha$ denotes the scheduling environment, $\beta$ describes the job and family characteristics and any restrictions, and $\gamma$ defines the objective function. This section focuses on relevant articles that deal with multiple parallel machines (i.e., $\alpha=P$). For a more comprehensive overview of single-machine p-batch scheduling, interested readers are referred to the recent survey by \citet{Fowler2022-SurveyP-Batching}. \nt{However, a notable mention should be made to the CP model by \citet{Malapert2012}, which solves instances of up to 50 compatible jobs in a single machine in just a few seconds thanks to their implementation of the pack and sequencing global constraints in Choco Solver \cite{ChocoSolver}.} 

For $\beta$, the focus is on articles that deal with multiple incompatible families (i.e., $P | p-batch, incompatible | \gamma$), although key articles with a single compatible family (i.e., $P | p-batch | \gamma$) are also considered, as the paper extends some of their concepts to the incompatible case. The review also includes some articles that consider non-identical job sizes ($s_j$) and release times ($r_j$). For $\gamma$, the reviewed objective functions primarily fall into three categories: (i) utilization objectives such as minimizing the makespan \nt{($C_{\max}$)}; (ii) cycle time objectives such as minimizing the total completion time (TCT) and its weighted counterpart (TWCT); and (iii) on-time delivery objectives such as minimizing earliness, lateness, tardiness, number of tardy jobs, including their weighted variants \cite{Fowler2022-SurveyP-Batching}. \rt{At bottleneck areas}, fab leaders often focus on \nt{two key aspects:} the cycle time of the jobs to increase the effective movement of products \cite{Wu2022-DiffusionFurnacesScheduling} \nt{and the utilization of the machines}.  \nt{This paper focuses} \rt{on the diffusion area using two objectives:} the TWCT, which specifically targets the cycle time performance \cite{Fowler2022-SurveyP-Batching}\nt{, and the makespan \nt{($C_{\max}$)}, which specifically targets the utilization performance \cite{Fowler2022-SurveyP-Batching}.}

Most of the articles that consider multiple parallel machines usually focus on a single family (i.e., $P | p-batch | \gamma$) \cite{Fowler2022-SurveyP-Batching}. This problem is NP-hard \cite{Lee1992, Fowler2022-SurveyP-Batching} and multiple heuristic algorithms \cite{Chung2009, Arroyo2017, Arroyo2019} as well as meta-heuristics \cite{Jia2018, Damodaran2012, Zhang2022} have been proposed. Some exact approaches include mixed-integer linear programming (MILP) models \cite{Trindade2018} and efficient arc-flow formulations \cite{Trindade2021, Trindade2020, Muter2020, Liu2022}. Recently, CP has been used for this problem in contexts different from SM, such as the composite manufacturing industry \cite{Tang2020}, and AM processes where parts with irregular shapes are 3D printed \cite{Nascimento2023, Nascimento2024, Cakici2024}. \nt{\citet{Tang2020} formulated a CP model and a logic-based benders decomposition (LBBD) for the composites manufacturing industry where the first stage is a compatible serial-batch of layers of raw materials and epoxy resin into mould tools, and the second stage is a compatible p-batch of the mould tools on autoclave ovens. Their CP model cannot find solutions to instances of more than 25 jobs but their LBBD does. In this paper we present four CP models that can find solutions to instances with up to 200 jobs. }\citet{Nascimento2023, Nascimento2024} presented an Assign-and-Schedule (AS) CP model that uses binary variables for the assignment of the \nt{irregular-shaped }parts to printing batches and CP scheduling logic to sequence the batches on the 3D printers. Their assignment logic already locates the parts inside the batch, so they do not explicitly include capacity constraints. However, they only consider compatible jobs (i.e., one family). This paper extends their formulation to handle multiple incompatible families, and include additional constraints to handle the batch size requirements.

In another recent development in the AM industry, \citet{Cakici2024} presented a synchronized CP model that replaces binary variables with interval variables, resulting in improved performance over the AS formulation. However, they consider only a single family of compatible jobs. This paper extends their model to handle multiple incompatible job families and impose additional constraints to accommodate batch size requirements. Furthermore, their model leverages optional interval variables in combination with a global synchronization constraint to ensure that jobs within the same batch start and finish together. 
\nt{A challenge faced in the AM industry is that the duration of the batches depends on the physical size of the parts being 3D printed in them. This results in only knowing the batch duration after the jobs are assigned to them. Their model defines a constraint that ensures the correct duration of the batches while simultaneously guaranteeing their presence when needed. In contrast, in the diffusion context studied in this paper, batches can be predefined to only group jobs of a specific job family, allowing to know \textit{a priori} their duration. Hence, such constraint that defines the batch duration is not necessary, but the presence of the correct batches is still required. This paper uses a different approach to do so and includes an explicit implication constraint instead.}

The literature that deals with multiple parallel machines and incompatible job families (i.e., $P | p-batch, incompatible | \gamma$) is rather scarce \cite{Fowler2022-SurveyP-Batching}. Since this problem is NP-hard \cite{Uzsoy1995}, the literature is dominated by heuristics and meta-heuristics \cite{Uzsoy1995, Koh2004, Kim2010, Klemmt2011, Li2022, Ji2023, Wu2022-DiffusionFurnacesScheduling}. However, exact methods have also been proposed, such as \nt{dynamic programming (DP)} \cite{Wu2022-DiffusionFurnacesScheduling} and MIPs \cite{Cakici2013, Ham2017-CP-p-batch-incompatible}. \citet{Cakici2013} proposed a time-indexed (TI) MIP model inspired by the truck delivery problem. They also presented two variable-neighborhood search (VNS) heuristics that outperform the MIP model. Later, \citet{Ham2017-CP-p-batch-incompatible} presented another MIP formulation based on positional assignment (PA) variables in the context of the diffusion area. They also proposed a CP model that outperforms both the PA and the TI MIP models, as well as the VNS heuristics. 

\nt{The core idea behind the model by \citet{Ham2017-CP-p-batch-incompatible} is also presented by \citet{Laborie2017}[Slide 129] and \citet{Laborie2018}. This core idea can be described in three parts. First, the usage of a state function that indicates the family being processed on a machine. This state function is a set of non-overlapping segments over which the function maintains a constant non-negative integer value (i.e., the family being processed). Second, the usage of a specialized $\AlwaysEqual$ global constraint that ensures that the state of a machine while processing a job is always equal to the job's family. Third, the usage of two optional alignment parameters in the $\AlwaysEqual$ constraint. These alignment parameters ensure that if a job is processed on a machine, its start and end times are \textit{aligned} with the start and end times of the time segment where the machine's state function takes the constant value. Because of this, the model presented by \citet{Ham2017-CP-p-batch-incompatible} is henceforward referred to as the \textit{Aligned} (A) model. This A model performs very well due to the propagation and pruning strategies behind the global constraint and alignment parameters in it. However, the state function, the $\AlwaysEqual$ constraint, and the two alignment parameters are only present in the commercial solver by IBM, ILOG CP Optimizer \cite{Laborie2018}.}

%%%%%%%%%%%%%%%%%%%%%%%%%%%%%%%%%%%%%%%%%%%%%%%%%%%%%%%%%%%%%%%%%%%%%%%%
% Contributions
%%%%%%%%%%%%%%%%%%%%%%%%%%%%%%%%%%%%%%%%%%%%%%%%%%%%%%%%%%%%%%%%%%%%%%%%

\section{Contributions} \label{sec: contributions}

As evidenced in the previous section, most of the existing CP models for p-batch only deal with a single compatible family \cite{Nascimento2023, Nascimento2024, Cakici2024}. \nt{The AS model by \citet{Cakici2024} relies on a constraint specifically designed to ensure the correct batch duration in the AM context, which does not directly translate to the diffusion problem \cite{Cakici2024}}. Moreover, \nt{the A model by \citet{Ham2017-CP-p-batch-incompatible} relies on a state function, a global constraint, and two extra alignment parameters that are only available in a commercial solver. Thus, the p-batch scheduling literature that considers multiple incompatible job families could benefit from new solver-independent CP models.}

This paper addresses these gaps in the literature by proposing \nt{four} new CP models that consider multiple incompatible job families and \nt{can be implemented in different commercial and open-source solvers}:  (i) \nt{A new \textit{Automaton} (AU) model that relies on automaton constraints, available in different open-source solvers, to handle the progression of the elapsed processing time of a batch. To the best of the authors' knowledge, this is the first CP model for p-batching that uses this type of constraints in the literature.} (ii) An \textit{Assign-and-Schedule} (AS) model inspired by \citet{Nascimento2023, Nascimento2024} that handles the assignment of jobs to batches with binary variables, and the sequencing of batches on machines using CP scheduling logic. (iii) A \textit{Synchronized} (S) model inspired by \citet{Cakici2024} that replaces the binary assignment with CP scheduling logic, \nt{and guarantees the presence of the required batches with an implication constraint}. (iv) A \textit{Redundant Synchronized} (RS) model that improves computational and quality performance \nt{through redundant variables and constraints.} The \nt{thorough} computational experiments \nt{on standard test cases compare these models using different CP solvers under different objectives and demonstrate the implementation flexibility of the proposed models.}

%%%%%%%%%%%%%%%%%%%%%%%%%%%%%%%%%%%%%%%%%%%%%%%%%%%%%%%%%%%%%%%%%%%%%%%%
% Problem description
%%%%%%%%%%%%%%%%%%%%%%%%%%%%%%%%%%%%%%%%%%%%%%%%%%%%%%%%%%%%%%%%%%%%%%%%

\section{Problem Description} \label{sec: problem description}

Let $\Jobs$ be the set of jobs, $\Families$ be the set of incompatible job families, and $\Machines$ be the set of machines. All jobs can be scheduled on all machines. Compatible jobs belong to the same family. Hence, let $f_j \in \Families$ be the family of job $j \in \Jobs$, and $\Jobs_f = \set{j \in \Jobs : f_j = f}$ be the subset of jobs that belong to family $f \in \Families$. All the jobs of the same family have the same processing time, thence, let $\ptime^f$ be the processing time of family $f$ and let $\ptime_j = \ptime^f$ be the processing time of job $j \in \Jobs_f$. Additionally, each job $j$ has a size $\size_j$ (i.e., number of parts), a weight $\weight_j$ (i.e., priority), and a release time $\releaseTime_j$. Each machine $m \in \Machines$ can simultaneously process multiple jobs of the same family $f \in \Families$ in batches whose \nt{maximum} size \nt{is} $u_f$. %, as evidenced in the instances of the SM testbed 2020 \cite{Kopp2020-FabTestbed}. 
Jobs in the same batch have to start and end together in a synchronized fashion, and jobs of different families cannot be processed together in the same batch. In the classification by \citet{Graham1979}, this problem is $P | s_j, r_j, p-batch, incompatible | \gamma$. \nt{For the objective $\gamma$, the goal is to minimize either the total weighted completion time (TWCT) or the makespan ($\makespan$) of the jobs.} 

%%%%%%%%%%%%%%%%%%%%%%%%%%%%%%%%%%%%%%%%%%%%%%%%%%%%%%%%%%%%%%%%%%%%%%%%
% CP overview
%%%%%%%%%%%%%%%%%%%%%%%%%%%%%%%%%%%%%%%%%%%%%%%%%%%%%%%%%%%%%%%%%%%%%%%%

\section{The Existing Aligned Model} \label{sec: existing models}

Model \ref{model: A} presents the A model proposed by \citet{Ham2017-CP-p-batch-incompatible} to schedule incompatible jobs in the diffusion area. Equation (\ref{eq: def interval job NS}) defines an interval variable $x_j$ of the form $[a, a+\ptime_j)$ (i.e., it has a duration of exactly $\ptime_j$ units of time), which represents job $j \in \Jobs$. \nt{Note that the start time of this interval must be inside of the scheduling horizon $\Horizon = \set{0, 1, \ldots, H}$, where $H = (\max_{j \in \Jobs} \releaseTime_j) + (\sum_{j \in \Jobs} \ptime_j) + 1$ is the horizon's limit.} Furthermore, this interval can only start after the job's release time.  Equation (\ref{eq: def interval job on machine NS}) defines an optional interval variable $x_\jm$, also of size $\ptime_j$, that represents the option of processing job $j$ on machine $m \in \Machines$. The presence of the interval $x_\jm$ is optional since it is allowed to take the value $\perp$, which indicates its absence in the solution. These optional interval variables allow the model to decide which machine should process the jobs. Equation (\ref{eq: state by family NS}) defines a function $\machineStateFamily_m$, which is the unique state of the machine $m$ at every point in time. Equation (\ref{eq: cumulative by family NS}) defines a cumulative function $\cumulLoad_{mf}$, which represents the load on machine $m$ of jobs of the family $f \in \Families$. It uses the $\pulse$ function, which receives two inputs: An interval $x_\jm$ and an integer value $\size_j$. This $\pulse$ function tells the cumulative function to increment its value in $\size_j$ units during the interval $x_\jm$, if present. Hence, the function $\cumulLoad_{mf}$ accumulates the sizes of overlapping jobs of family $f \in \Families$ on machine $m$ at every point in time.

\begin{model}[!t]
\caption{Aligned model by \citet{Ham2017-CP-p-batch-incompatible}} \label{model: A}
\renewcommand{\arraystretch}{1.3}

\begin{tabularx}{\linewidth}{@{}R@{}LK@{}R@{}}
    \multicolumn{4}{@{}l}{Variables:} \\
    x_j & \in \{[a,a + \ptime_j): \releaseTime_j \leq a \in \Horizon \}, & \forall ~j \in \Jobs; & \AddLabel{eq: def interval job NS} \\
    x_\jm & \in \{[a,a + \ptime_j) : a \in \Horizon \} \cup \{\perp\}, & \forall ~ j \in \Jobs, m \in \Machines; & \AddLabel{eq: def interval job on machine NS}\\
    \multicolumn{4}{@{}l}{Functions:} \\
    \machineStateFamily_m &: \StateFunction & \forall ~m \in \Machines; & \AddLabel{eq: state by family NS}\\
    \cumulLoad_{mf} &= \sum_{j \in \Jobs_f} \pulse \pp{x_\jm, \size_j}, & \forall ~ m \in \Machines, f \in \Families. & \AddLabel{eq: cumulative by family NS}\\
\end{tabularx}

\begin{tabularx}{\linewidth}{@{}cXR@{}LR@{}}
% Row 1: Objective function
\multicolumn{5}{@{}l}{Formulation:} \\
    ~ & \multicolumn{3}{C}{\text{minimize~} \sum_{j \in \Jobs} \weight_j \cdot \EndOf\pp{x_j},} & \AddLabel{eq: objective function twct NS} \\
% Subject to
\multicolumn{5}{l}{subject to,} \\
% One job on one machine
    ~ & \multicolumn{3}{L}{\forall ~j \in \Jobs:} & \multirow{2}{*}{\AddLabel{eq: job on one machine NS}} \\
    \multicolumn{2}{R@{}}{\phantom{\texttt{alternat}}\Alternative} & \multicolumn{2}{@{}K}{\pp{x_j, \set{x_\jm}_{m \in \Machines}};} & \\
    % ~ & ~ & \hfil \Alternative & \pp{x_j, \set{x_\jm}_{m \in \Machines}};  & \\
% AlwaysEqual by family
    ~ & \multicolumn{3}{L}{\forall ~m \in \Machines, j \in \Jobs:} & \multirow{2}{*}{\AddLabel{eq: always equal by family NS}} \\
    \multicolumn{2}{R@{}}{\AlwaysEqual} & \multicolumn{2}{@{}K}{\pp{\machineStateFamily_m, x_\jm, f_j, \nt{\texttt{True}, \texttt{True}}};} & \\
    % ~ & ~ & \AlwaysEqual & \pp{\machineStateFamily_m, x_\jm, f_j}; & \\
% AlwaysIn by family
    % ~ & \multicolumn{3}{L}{\forall ~ m \in \Machines, f \in \Families:} & \multirow{2}{*}{\AddLabel{eq: cumulative bound by family NS}} \\
    % \multicolumn{2}{R@{}}{\cumulLoad_{mf}} & \multicolumn{2}{@{}K}{ \leq u_f.} & \\
    ~ & \multicolumn{3}{L}{\forall ~ m \in \Machines, f \in \Families: \quad \cumulLoad_{mf} \leq u_f.} & \AddLabel{eq: cumulative bound by family NS} \\
    % ~ & ~ & \AlwaysIn & \pp{\cumulLoad_{mf}, x_\jm, l_f, u_f}. & \\
    % \bottomrule
\end{tabularx}
\end{model}

Objective function (\ref{eq: objective function twct NS}) minimizes the TWCT. Constraints (\ref{eq: job on one machine NS}) use the $\Alternative\nt{(v, V)}$ global constraint, which receives an interval variable $v$, and a set of optional interval variables $V$. This global constraint ensures that, if the interval variable $v$ is present, then only one of the optional interval variables in $V$ gets selected to be present, and $v$ gets synchronized with the selected interval in $V$ \cite{ibm2023cpoptimizer}. Hence, constraints (\ref{eq: job on one machine NS}) enforce each job to be processed on exactly one machine. Constraints (\ref{eq: always equal by family NS}) use the global constraint $\AlwaysEqual\nt{(h, v, k, align\_s, align\_e)}$, which receives a state function $h$, an interval variable $v$, an integer value \nt{${k}$, and two optional boolean parameters $align\_s$ and $align\_e$}. This global constraint ensures that the state function $h=k$ during the interval $v$, if present \cite{ibm2023cpoptimizer}. \nt{Besides, if the two boolean parameters are set to \texttt{True}, the start and end times of the interval $v$ are aligned with the start and end times of the time segment where the state function $h$ takes the value of $k$}. Hence, constraints (\ref{eq: always equal by family NS}) ensure that jobs of only one family are processed at any time on the machines \nt{in batches, and that jobs in the same batch start and end together}. Finally, constraints (\ref{eq: cumulative bound by family NS}) \nt{impose an upper bound on the cumulative functions that captures the load on the machines by family, ensuring the maximum batch sizes.}

\nt{
To minimize the makespan, objective function \eqref{eq: objective function twct NS} should be replaced by objective function \eqref{eq: objective function cmax NS}, which minimizes the maximum end time of the jobs. Although this nonlinear objective function should be linearized in a linear programming (LP) model, a CP solver is able to handle it as presented.
\begin{equation}
    \text{minimize~~} \max_{j \in \Jobs} ~ \EndOf \pp{x_j} \label{eq: objective function cmax NS}
\end{equation}

Consider the scenario in Table \ref{tab: Simple example} as an illustrative example with four jobs, one family, and one machine. Assume that the size of each batch must be at most $u_1=100$. Figure \ref{fig: A model solutions} shows the Gantt chart of different solutions of the A model in different settings. Figure \ref{fig: aligned solution} shows the solution with these alignment parameters set to \texttt{True} and minimizing the TWCT. Note how the machine state is divided in two segments and the start and end times of the jobs are aligned with the start and end times of their corresponding state segment. Figure \ref{fig: aligned solution cmax} shows the aligned solution minimizing the makespan. In constrast, Figure \ref{fig: non-synchronized solution} shows the solution with either objective and the alignment parameters set to $\texttt{False}$, which allows jobs to start mid-process of other jobs, making this unaligned solution infeasible.

\begin{table}[!t]
    \caption{Information of the illustrative example}
    \label{tab: Simple example}
    \centering
    \begin{tabular}{ccccccc}
        \toprule
        \makecell{Job \\ $j$} & \makecell{Size \\ $\size_j$} & \makecell{Weight \\ $\weight_j$} & \makecell{Release\\ time $\releaseTime_j$}  & \makecell{Family \\ $f_j$} & \makecell{Processing \\ time $\ptime_j$} \\
        \midrule
        1 & 25 & 10 & 3  & 1 & 10 \\
        2 & 25 & 10 & 11  & 1 & 10 \\
        3 & 25 & 20 & 5  & 1 & 10 \\
        4 & 25 & 40 & 12  & 1 & 10 \\
        \bottomrule
    \end{tabular}
\end{table}

\begin{figure}[!t]
    \centering
    \begin{subfigure}[b]{\linewidth}
        \centering
        \includegraphics[width=\textwidth]{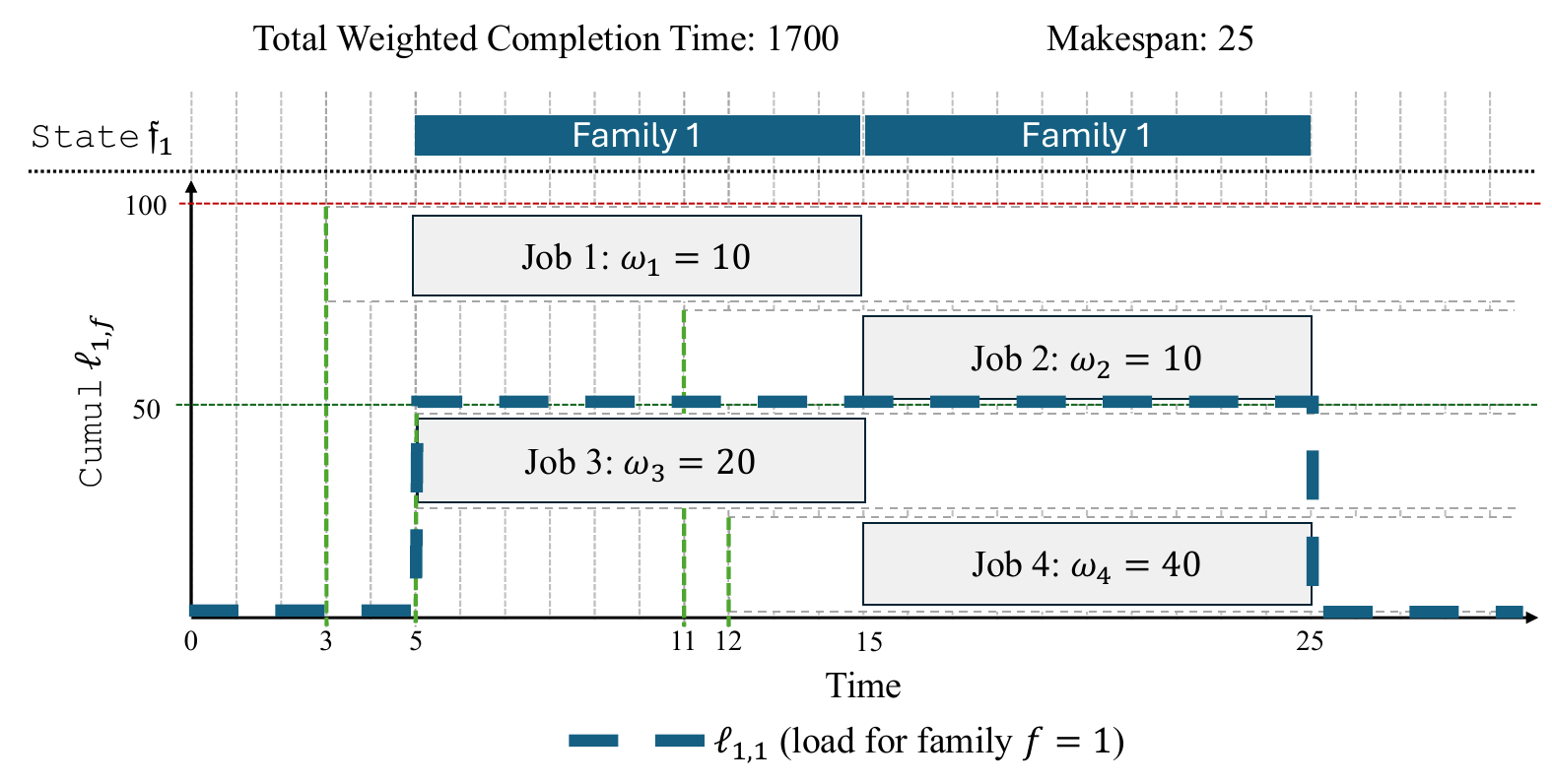}
        \vspace{-1.5em}
        \caption{TWCT and alignment parameters set to \texttt{True}}
        \label{fig: aligned solution}
    \end{subfigure}
    \\
    \begin{subfigure}[b]{\linewidth}
        \centering
        \includegraphics[width=\textwidth]{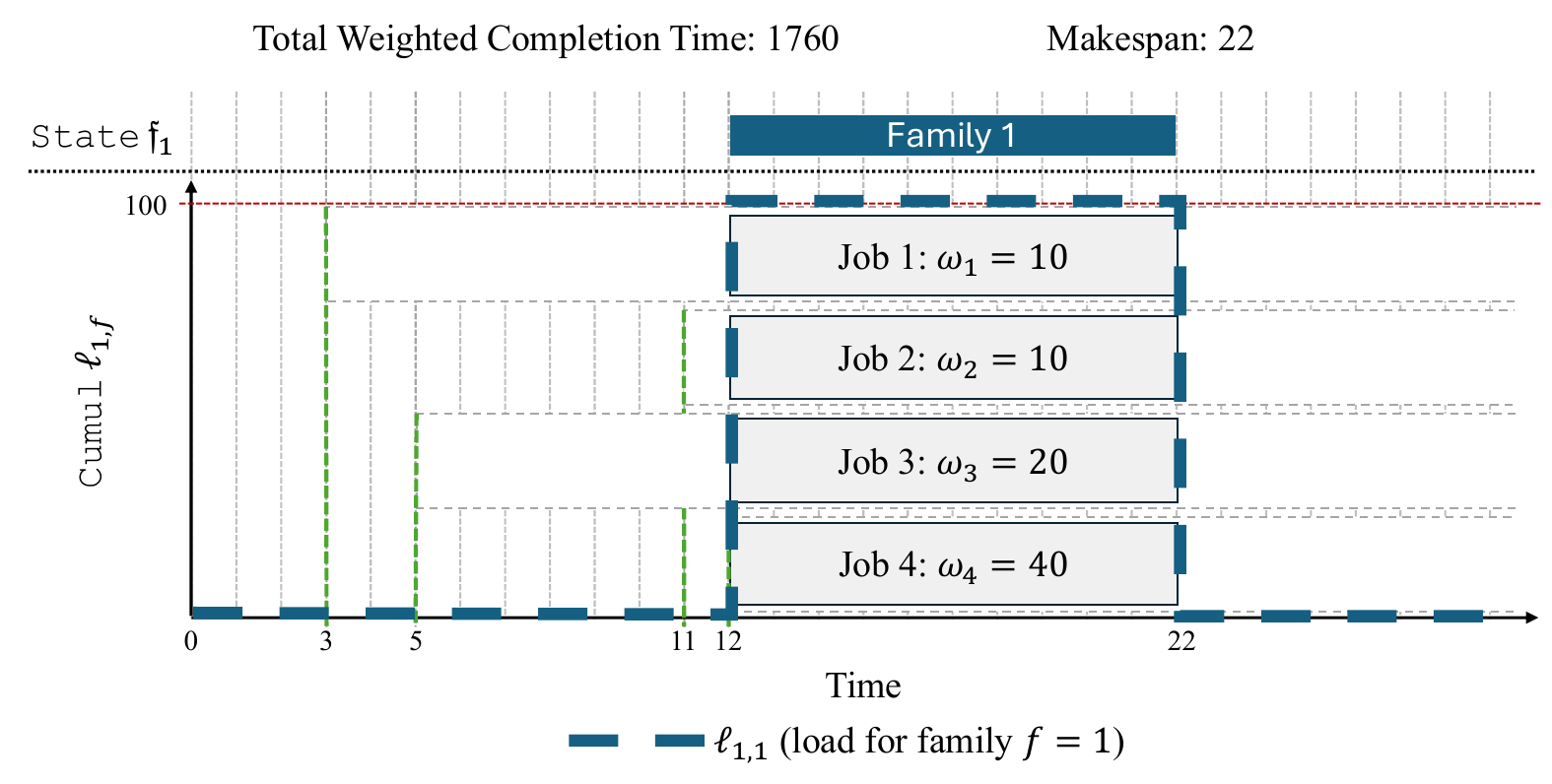}
        \vspace{-1.5em}
        \caption{$\makespan$ and alignment parameters set to \texttt{True}}
        \label{fig: aligned solution cmax}
    \end{subfigure}
    \\
    \begin{subfigure}[b]{\linewidth}
        \centering
        \includegraphics[width=\textwidth]{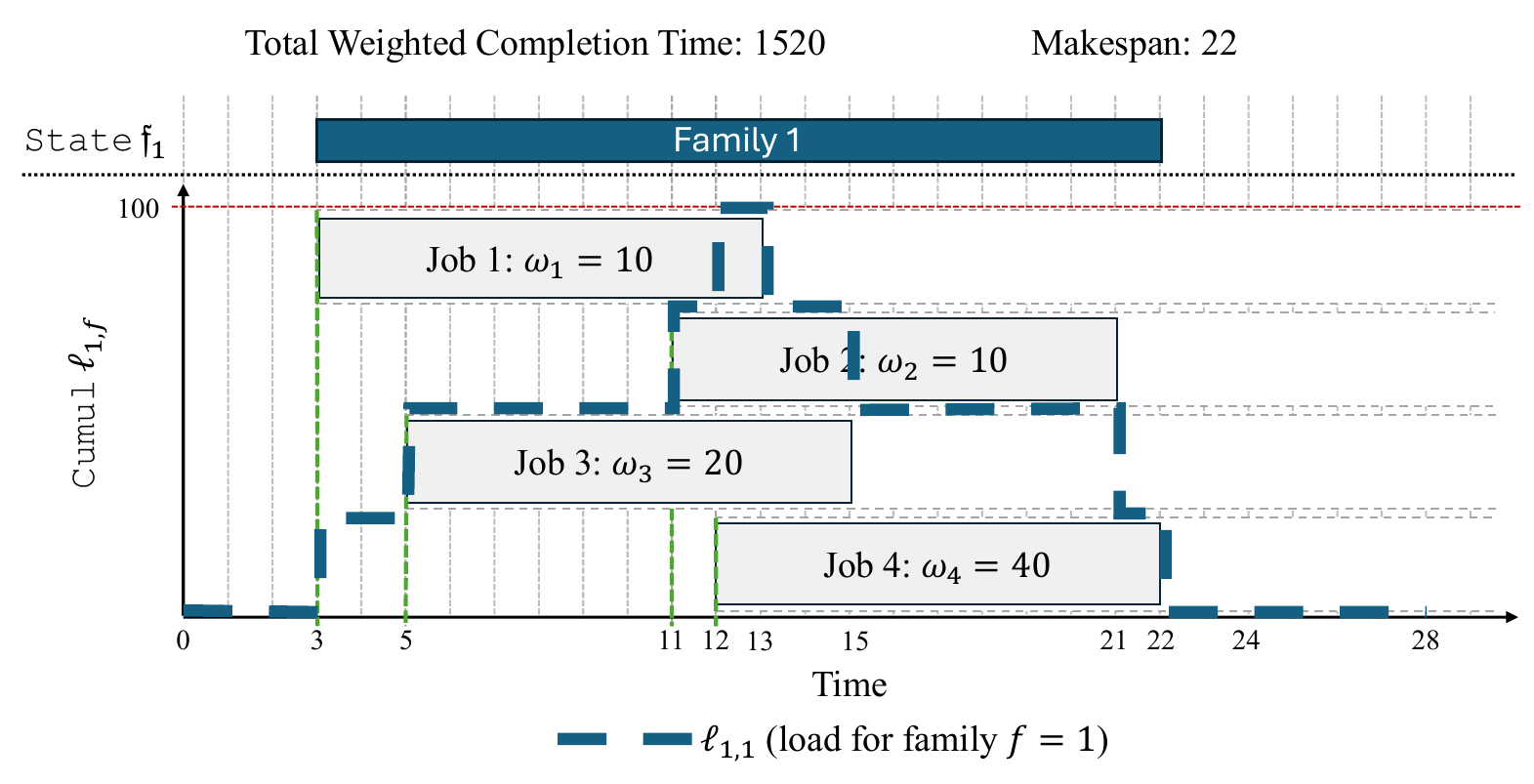}
        \vspace{-1.5em}
        \caption{Alignment parameters set to \texttt{False}}
        \label{fig: non-synchronized solution}
    \end{subfigure}
    \caption{Optimal solutions in different settings}
    \label{fig: A model solutions}
\end{figure}

Although the A model is the current state-of-the-art CP model for p-batch scheduling with incompatible job families, the state function \eqref{eq: state by family NS} and the global constraint \eqref{eq: always equal by family NS} are only available in the commercial solver IBM ILOG CP Optimizer, making it solver-dependent. Similar to how an LP model is independent of the solver implementation, the scheduling literature could benefit from new CP models that can be implemented in other commercial and open-source solvers.
}
%%%%%%%%%%%%%%%%%%%%%%%%%%%%%%%%%%%%%%%%%%%%%%%%%%%%%%%%%%%%%%%%%%%%%%%%
% Proposed models
%%%%%%%%%%%%%%%%%%%%%%%%%%%%%%%%%%%%%%%%%%%%%%%%%%%%%%%%%%%%%%%%%%%%%%%%

\section{The Proposed CP Models} \label{sec: proposed models}
 \nt{
This section presents the four proposed CP models that can be implemented in other commercial and open-source solvers. 

%%%%%%%%%%%%%%%%%%%%%%%%%%%%%%%%%%%%%%%%%%%%%%%%%%%%%%%%%%%%%%%%%%%%%%%%
% Automaton model
%%%%%%%%%%%%%%%%%%%%%%%%%%%%%%%%%%%%%%%%%%%%%%%%%%%%%%%%%%%%%%%%%%%%%%%%

\subsection{The Automaton model} \label{sec: automaton model}

The Automaton (AU) model derives its name from the $\Automaton$ global constraint it relies on, which is available in different open-source CP solvers \cite{cpsatlp, ChocoSolver, MiniZinc-Nethercote2007}. Automaton constraints model finite state machines, enabling the representation of feasible transitions between states. In the AU model, these finite states represent the elapsed time when processing a batch of each family. Consider the automaton depicted in Figure \ref{fig:automaton}, where nodes are states and arcs are transition possibilities. The descriptions in each state help understand the figure and the numbers in parenthesis ultimately identify them. Each arc is identified by a triplet $(i,k,\tau)$, where $i$ is the initial state, $k$ is the end state, and $\tau$ is the required transition value to trigger such transition from $i$ to $k$. For example, if a machine is empty and receives a transition value of 0, then the machine remains empty. But if it receives a transition value of 1, it means that a new batch of family 1 starts on the machine. From this point, the only possible transition values that the automaton can receive are the \rt{ones} necessary to progress the states until completing the processing time $\ptime^1$ of the batch. Once the batch is completed, three possible transitions can be made: either the machine becomes empty (0), or a new batch of family 1 starts (1), or a new batch of family 2 starts ($\ptime^1 + 1$). From that point, a new sequence of transition values must be given such that the state transitions respect the automaton structure.

\begin{figure}
    \centering
    \includegraphics[width=\linewidth]{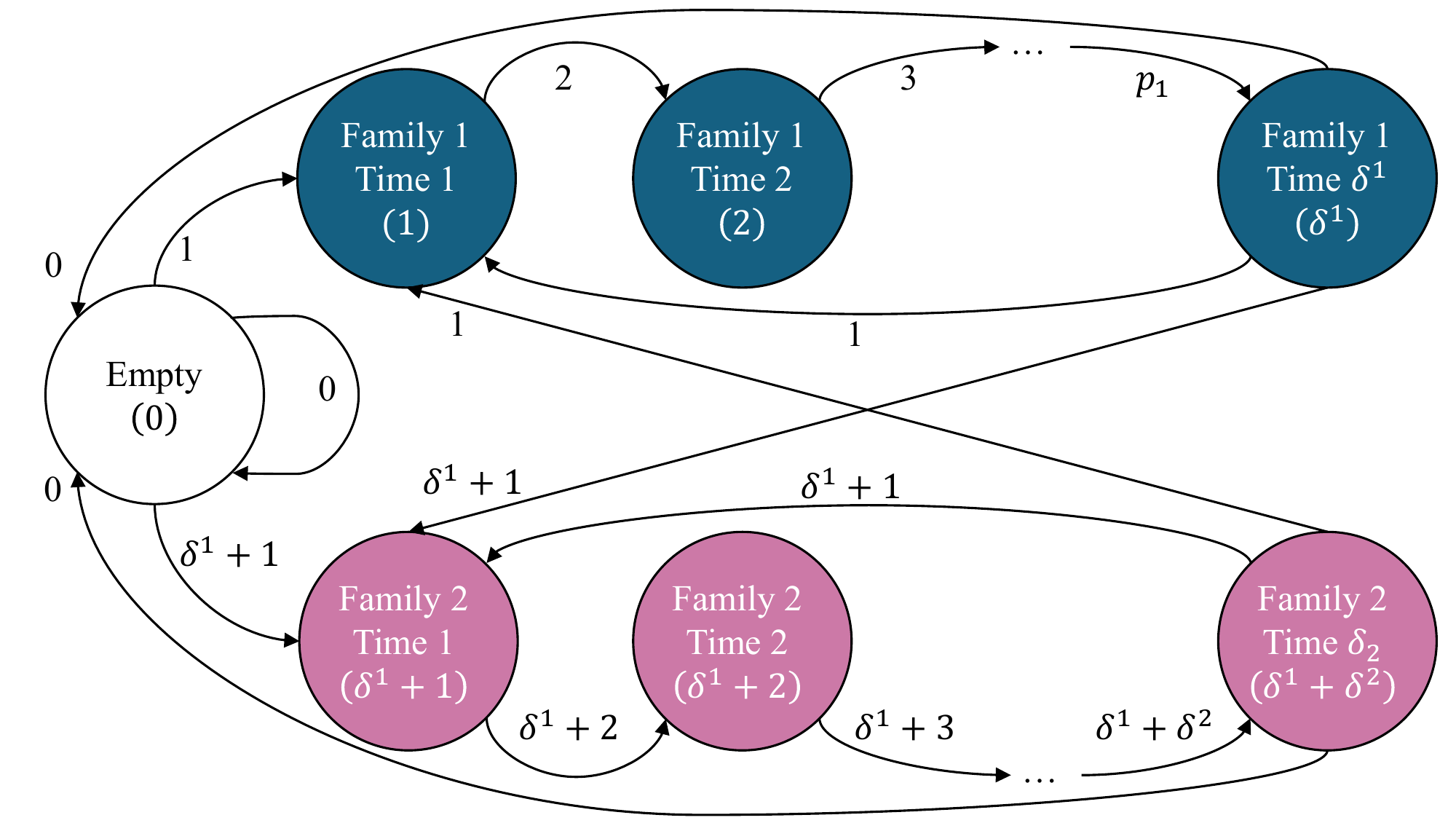}
    \caption{Example automaton with two job families}
    \label{fig:automaton}
\end{figure}

Let $\Graph = (\Nodes, \Arcs)$ be the graph describing the automaton. Let $\Nodes = \set{i : 0 \leq i \leq \sum_{f \in \Families} \ptime^f}$ be the nodes that represent finite states, where $0$ represents the empty state. Note that each family has a sequence of states. Let $\inistate_f \in \Nodes$ and $\finalstate_f \in \Nodes$ be the initial and end states in the sequence of states of family $f \in \Families$, respectively. By setting $\ptime^0 = 0$, then $\inistate_f = \ptime^{f-1} + 1$ and $\finalstate_f = \ptime^{f-1} + \ptime^f$. In this way, let $\InitialSatets = \set{\inistate_f : f \in \Families}$ and $\FinalStates = \set{\finalstate_f : f \in \Families}$ be the sets of all initial and final states of the families, respectively. Let $\Arcs = \set{(i,k,\tau): i,k \in \Nodes, \tau \in \integers^1_+}$ be the set of arcs that identify the possible transitions between states and the transition values that trigger them. This set of transition triplets can be further divided in five subsets, i.e., $\Arcs = \Arcs_0 \cup \Arcs_1 \cup \Arcs_2 \cup \Arcs_3 \cup \Arcs_4$. Let $\Arcs_0 = \set{(0,0,0)}$ be the transition triplet that allows an empty machine remain empty. Let $\Arcs_1 = \set{(0, \inistate_f, \inistate_f) : f \in \Families}$ be the transition triplets that allow an empty machine start a new batch of any family. Let $\Arcs_2 = \set{(i, i+1, i+1): \inistate_f \leq i < \finalstate_f, f \in \Families}$ be the transition triplets that allow the progression throughout the processing time of a batch of any family. Let $\Arcs_3 = \set{(\finalstate_f, 0, 0): f \in \Families}$ be the transition triplets that allow a machine that completes the processing time of a batch change to the empty state. Finally, let $\Arcs_4 = \set{(\finalstate_f, \inistate_g, \inistate_g): f,g \in \Families}$ be the transition triplets that allow a machine that completes the processing time of a batch start the processing of a new batch, even of different family.

Model \ref{model: AU} presents the AU model using the syntax of CP-SAT \cite{cpsatlp}, an open-source solver in Google OR-Tools \cite{ortools}. For the variables of this model, equation \eqref{eq: def start times AU} defines the start time of job $j \in \Jobs$ on machine $m \in \Machines$. Equation \eqref{eq: def presences AU} defines a boolean variable $p_\jm$ that takes the value of 1 if job $j$ is processed on machine $m$ or 0 if not. Equation \eqref{eq: def interval job on machine AU} defines the interval variable $x_\jm$ with presence literal $p_\jm$, which represents job $j$ being processed on machine $m$. Equation \eqref{eq: def modified start time AU} defines a modified start time $\hat{s}_\jm$ of the job $j$ on machine $m$ which is the same start time $s_\jm$ if the job is processed on the machine (i.e., $p_\jm = 1$), or the horizon's limit $H$ if not. Equation \eqref{eq: def transition values AU} defines the transition values of machine $m$'s automaton at each time step $t$ of the scheduling horizon $\Horizon$. Finally, equation \eqref{eq: def modified transition AU} defines the transition value $\hat{\tau}_\jm$ at the modified start time $\hat{s}_\jm$ on machine $m$, which is the initial state $\inistate_{f_j}$ if job $j$ is processed on machine $m$, or the empty state otherwise.

\begin{model}[!t]
\caption{Automaton model} \label{model: AU}
\renewcommand{\arraystretch}{1.3}

\begin{tabularx}{\linewidth}{@{}R@{}LK@{}R@{}}
    \multicolumn{4}{@{}l}{Variables:} \\
    s_\jm & \in \set{a \in \Horizon: \releaseTime_j \leq a}, & \forall ~j \in \Jobs, m \in \Machines; & \AddLabel{eq: def start times AU}\\
    p_\jm & \in \set{ 0, 1},  & \forall ~j \in \Jobs, m \in \Machines; & \AddLabel{eq: def presences AU}\\
    x_\jm & = [s_\jm, s_\jm + \ptime_j) \text{ with } p_\jm,  & \forall ~ j \in \Jobs, m \in \Machines; & \AddLabel{eq: def interval job on machine AU}\\
    \hat{s}_\jm & = s_\jm \cdot p_\jm + H \cdot (1 - p_\jm), & \forall ~j \in \Jobs, m \in \Machines; & \AddLabel{eq: def modified start time AU}\\
    \tau_{mt} & \in \Nodes, & \forall ~m \in \Machines, t \in \Horizon; & \AddLabel{eq: def transition values AU}\\
    \hat{\tau}_\jm & = \inistate_{f_j} \cdot p_\jm \in \InitialSatets \cup \{0\}, & \forall ~j \in \Jobs, m \in \Machines; & \AddLabel{eq: def modified transition AU}\\
\end{tabularx}

\begin{tabularx}{\linewidth}{@{}cXR@{}LR@{}}
% Row 1: Objective function
\multicolumn{5}{@{}l}{Formulation:} \\
    ~ & \multicolumn{3}{C}{\text{minimize~} \sum_{j \in \Jobs} \sum_{m \in \Machines} \weight_j \cdot p_\jm \cdot (s_\jm + \ptime_j),} & \AddLabel{eq: objective function twct AU} \\
% Subject to
\multicolumn{5}{l}{subject to,} \\
% One job on one machine
    ~ & \multicolumn{3}{L}{\forall ~j \in \Jobs:} & \multirow{2}{*}{\AddLabel{eq: exactly one machine for job AU}} \\
    \multicolumn{2}{R@{}}{\phantom{\texttt{aout}}\ExactlyOne} & \multicolumn{2}{@{}K}{\pp{\{p_\jm\}_{m \in \Machines}};} & \\
    % ~ & ~ & \hfil \Alternative & \pp{x_j, \set{x_\jm}_{m \in \Machines}};  & \\
% Cumulative 
    ~ & \multicolumn{3}{L}{\forall ~m \in \Machines, f \in \Families:} & \multirow{2}{*}{\AddLabel{eq: cumulative AU}}\\
    \multicolumn{2}{R@{}}{\texttt{cumulative}} & \multicolumn{2}{@{}K}{\pp{\{x_\jm\}_{j \in \Jobs_f}, \{\size_j\}_{j \in \Jobs_f}, u_f}; } & \\
% Element
    ~ & \multicolumn{3}{L}{\forall ~j \in \Jobs, m \in \Machines:} & \multirow{2}{*}{\AddLabel{eq: element AU}}\\
    \multicolumn{2}{R@{}}{\Element} & \multicolumn{2}{@{}K}{\pp{ \hat{s}_\jm, \{\tau_{mt} \}_{t \in \Horizon}, \hat{\tau}_\jm};} & \\
% Automaton
    ~ & \multicolumn{3}{L}{\forall ~m \in \Machines, f \in \Families:} & \multirow{2}{*}{\AddLabel{eq: Automaton AU}}\\
    \multicolumn{2}{R@{}}{\Automaton} & \multicolumn{2}{@{}K}{\pp{ \{ \tau_{mt} \}_{t \in \Horizon}, 0, \FinalStates \cup \{0 \}, \Arcs }.} & \\
\end{tabularx}
\end{model}

Objective function \eqref{eq: objective function twct AU} minimizes the TWCT. Constraints \eqref{eq: exactly one machine for job AU} use the global constraint $\ExactlyOne(P)$, which receives a set of boolean variables $P$ and gives the value of 1 to only one of them. Hence, constraints \eqref{eq: exactly one machine for job AU} ensure that each job is processed on exactly one machine. Constraints \eqref{eq: cumulative AU} use the global constraint $\texttt{cumulative}(V, D, u)$, which receives a set of interval variables $V$, their demands $D$, and the capacity $u$ of a resource. This constraint ensures that the demands of overlapping intervals respect the resource capacity. Hence, constraints \eqref{eq: cumulative AU} ensure the maximum batch size of every family. Constraints \eqref{eq: element AU} use the global constraint $\Element(t, T, \tau)$, which receives an integer decision variable $t$, a sequence of integer decision variables  $T$, and an integer decision variable $\tau$. This constraint ensures that the $t$\textsuperscript{th} element of the sequence $T$ takes the value of $\tau$ (i.e., $T_t = \tau$). Hence, constraints \eqref{eq: element AU} ensure that if job $j$ is processed on the machine $m$, then the machine's transition value at time $s_\jm$ is the initial automaton state $\inistate_{f_j}$, (i.e., $\tau_{m, s_\jm} = \inistate_{f_j}$); but if the job is not processed on such machine, then the transition value at the end of the scheduling horizon leaves the machine empty (i.e., $\tau_{m, H} = 0$). Thus, constraints \eqref{eq: element AU} guarantee that every time a job starts on a machine, the right transition value is provided to the machine's automaton. Finally, constraints \eqref{eq: Automaton AU} use the global constraint $\Automaton(T, \inistate, \FinalStates, \Arcs)$ which receives a sequence of transition variables $T$, an initial state $\inistate$, a set of possible end states $\FinalStates$, and a set defining the transition triples $\Arcs$ of the automaton. This constraint ensures that starting from the initial state $\inistate$, the transition variables $T$ take values that respect the automaton's structure and end in one of the final states $\FinalStates$. Thence, constraints \eqref{eq: Automaton AU} ensure that the machines start empty and transition their states respecting the automaton's structure, until reaching one of the final states $\FinalStates$ at the end of the scheduling horizon, or remain empty. Since there is only one possible transition after reaching the initial state $\inistate_f$ of any family $f$, this constraint will ensure that jobs of the same family processed on each machine don't interrupt any other job already started. Therefore, due to the automaton's structure, all the jobs in the same batch start and end together. 
% Note that constraints \eqref{eq: element AU} achieve a reified version of the element constraint that only takes place when a job is actually processed on each machine. Nonetheless, this reification is not explicit. Instead, it is achieved indirectly through appropriately setting the values of the modified start times $\hat{s}_\jm$ and transition values $\hat{\tau}_\jm$.

To minimize the makespan, objective function \eqref{eq: objective function twct AU} should be replaced by objective function \eqref{eq: objective function cmax AU}, which minimizes the maximum end time of the jobs.
\begin{equation}
    \text{minimize~~} \max_{j \in \Jobs, m \in \Machines} ~ p_\jm \cdot (s_\jm + \ptime_j) \label{eq: objective function cmax AU}
\end{equation}

Figure \ref{fig: AU solution} presents the optimal solution of the AU model minimizing the TWCT. Empty states don't have the id. The initial state of the family is in blue and the final state is in magenta. The transition values are below the arcs. Note that in each batch, both jobs send the same transition value to the automaton, which aligns them.

\begin{figure}[!t]
    \centering
    \includegraphics[width=\linewidth]{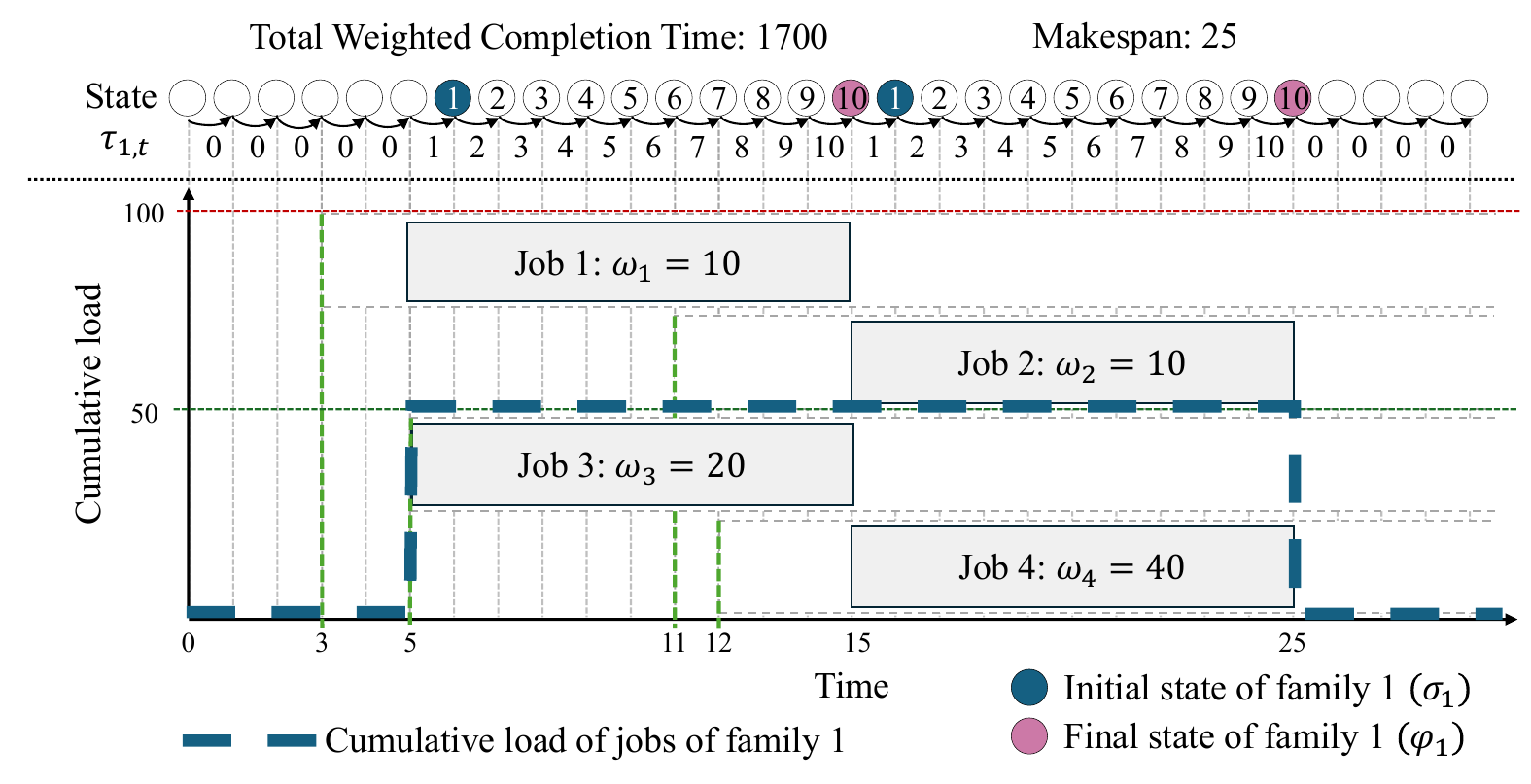}
    \caption{AU optimal solution minimizing TWCT}
    \label{fig: AU solution}
\end{figure}
}
%%%%%%%%%%%%%%%%%%%%%%%%%%%%%%%%%%%%%%%%%%%%%%%%%%%%%%%%%%%%%%%%%%%%%%%%
% Assign-and-Schedule model
%%%%%%%%%%%%%%%%%%%%%%%%%%%%%%%%%%%%%%%%%%%%%%%%%%%%%%%%%%%%%%%%%%%%%%%%
\subsection{The Assign-and-Schedule Model} \label{sec: assign and schedule model}

The following three models make explicit use of the ordered set of possible batches $\Batches = \set{1,2, 3\ldots}$. The family $f^b \in \Families$ of each possible batch $b \in \Batches$ is defined \textit{a priori}. In addition, the model only allows jobs $\Jobs_{f^b}$ of family $f^b$ to be processed on batch $b$. Hence, this set can be further partitioned into the mutually exclusive sets $\Batches_f = \set{b \in \Batches : f^b = f} \subset \Batches$ of possible batches for each family $f \in \Families$. 
In this way, the first $|\Batches_1|$ elements of $\Batches$ correspond to the possible batches where jobs of family $1$ can be processed; the next $|\Batches_2|$ elements to the possible batches for jobs of the family $2$, and so on. 
The number of possible batches of each family is upper bounded by the number of jobs that belong to that family, hence, we set $|\Batches_f|=|\Jobs_f|$.

\begin{model}[!t]
\caption{The Assign-and-Schedule Model Inspired by \cite{Nascimento2023}} \label{model: AS}
\renewcommand{\arraystretch}{1.3}

\begin{tabularx}{\linewidth}{@{}R@{}LK@{}R@{}}
    \multicolumn{4}{@{}l}{Variables:} \\
    z_\jb & \in \set{0,1}, & \forall ~ j \in \Jobs, b \in \Batches_{f_j}; & \AddLabel{eq: def binary job in batch AS}\\
    y_b & \in \{[a,a + \ptime^{f^b}): a \in \Horizon\} \cup \{\perp\}, & \forall ~b \in \Batches; & \AddLabel{eq: def interval batch AS} \\
    y_\bm & \in \{[a,a + \ptime^{f^b}) : a \in \Horizon \} \cup \{\perp\}, & \forall ~ b \in \Batches, m \in \Machines; & \AddLabel{eq: def interval batch on machine AS}\\
    \varphi_m & \in \Permutation\pp{\set{y_\bm}_{b \in \Batches}}, & \forall ~m \in \Machines. & \AddLabel{eq: def sequence var AS}
\end{tabularx}
\noindent
\begin{tabularx}{\linewidth}{@{}cXR@{~}LR@{}}
\multicolumn{4}{@{}l}{Formulation:} \\
% Row 1: Objective function
    ~ & \multicolumn{3}{C}{\text{minimize~} \sum_{j \in \Jobs} \weight_j \cdot \sum_{b \in \Batches_{f_j}} \EndOf\pp{y_b} \cdot z_\jb,} & \AddLabel{eq: objective function twct AS} \\
% Subject to
\multicolumn{5}{l}{subject to,} \\
% Assignment
    ~ & \multicolumn{3}{L}{\forall ~ j \in \Jobs:} & \multirow{2}{*}{\AddLabel{eq: binary assignment}} \\
    ~ & ~ & \sum_{b \in \Batches_{f_j}} z_\jb ~ & = 1;  & \\
% Activation
    ~ & \multicolumn{3}{L}{\forall ~b \in \Batches:} &  \\
    ~ & ~ & \sum_{j \in \Jobs_{f^b}} z_\jb ~ & \geq \PresenceOf\pp{y_b}; & \AddLabel{eq: binary activation lb}\\
    ~ & ~ & \sum_{j \in \Jobs_{f^b}} z_\jb ~ & \leq |\Jobs_{f^b}| \cdot \PresenceOf\pp{y_b}; & \AddLabel{eq: binary activation ub}\\
% Capacity
    ~ & \multicolumn{3}{L}{\forall ~b \in \Batches:} &  \\
    % ~ & ~ & \sum_{j \in \Jobs_{f^b}} \size_j \cdot z_\jb ~ & \geq l_{f^b} \cdot \PresenceOf\pp{y_b}; & \AddLabel{eq: capacity lb}\\
    ~ & ~ & \sum_{j \in \Jobs_{f^b}} \size_j \cdot z_\jb ~ & \leq u_{f^b} \cdot  \PresenceOf\pp{y_b}; & \AddLabel{eq: capacity ub}\\
% Release time
    ~ & \multicolumn{3}{L}{\forall ~ b \in \Batches, j \in \Jobs_{f^b}: } & \multirow{2}{*}{\AddLabel{eq: release time}} \\
    ~ & ~ & \StartOf\pp{y_b} & \geq \releaseTime_j \cdot z_\jb; & \\
% If batch => batch on one machine
    ~ & \multicolumn{3}{L}{\forall ~b \in \Batches: } & \multirow{2}{*}{\AddLabel{eq: each batch on one machine AS}} \\
    ~ & ~ & \Alternative & \pp{y_b , \set{y_\bm}_{m \in \Machines}}; & \\
% No overlap of batches
    ~ & \multicolumn{3}{L}{\forall ~ m \in \Machines: } & \multirow{2}{*}{\AddLabel{eq: no overlapping AS}} \\
    ~ & ~ & \NoOverlap & \pp{\varphi_m}. & \\
% \bottomrule
\end{tabularx}
\end{model}

Model \ref{model: AS} presents the Assign-and-Schedule (AS) model \nt{using the syntax of the commercial solver CP Optimizer. However, the same model can be implemented in open-source solvers}. It is based on the CP model by \citet{Nascimento2023} for AM processes, which deals with a single compatible family and does not include capacity constraints. This paper extends their formulation to handle multiple incompatible families and include the necessary constraints to ensure the \nt{maximum} batch sizes. The model name originates from its clear distinction between assignment and scheduling decisions. While the job assignment to batches is handled with binary variables, the batch sequencing is handled with CP scheduling logic.

Equation (\ref{eq: def binary job in batch AS}) defines a binary variable $z_\jb$ that takes the value of 1 if job $j \in \Jobs$ is assigned to batch $b \in \Batches_{f_j}$, or 0 otherwise. Equation (\ref{eq: def interval batch AS}) defines an optional interval variable $y_b$ of size $\ptime^{f^b}$ that represents batch $b \in \Batches$. It is optional because a batch with no jobs assigned should not be present in the solution. Equation (\ref{eq: def interval batch on machine AS}) defines an optional interval variable $y_\bm$, also of size $\ptime^{f^b}$, that represents batch $b$ being processed on machine $m \in \Machines$. Equation (\ref{eq: def sequence var AS}) defines an \textit{interval sequence} variable $\varphi_m$, which represents the sequence of batches scheduled on machine $m$. This sequence variable can be any permutation of the batch intervals $\set{y_\bm}_{b \in \Batches}$ on the machine $m$ present in the solution.

Objective function (\ref{eq: objective function twct AS}) minimizes the TWCT. Constraints (\ref{eq: binary assignment}) assign each job to exactly one possible batch. Constraints (\ref{eq: binary activation lb}) deactivate the presence of the batches if no jobs are assigned to them. Constraints (\ref{eq: binary activation ub}) activate the presence of the batches if at least one job is assigned to them. Constraints (\ref{eq: capacity ub}) ensure that if a batch is used, its size \nt{respects the maximum batch capacity}. Constraints (\ref{eq: release time}) guarantee that batches are scheduled once all the jobs in them are available. Constraints (\ref{eq: each batch on one machine AS}) schedule each batch on exactly one machine, if present. Finally, constraints (\ref{eq: no overlapping AS}) use the global constraint $\NoOverlap\rt{(\varphi)}$, which \rt{receives a sequence variable $\varphi$. This constraint} ensures that the present interval variables in the provided sequence not overlap. Hence, constraints (\ref{eq: no overlapping AS}) guarantee that only one batch is processed at a time on each machine.

\nt{To minimize the makespan, objective function \eqref{eq: objective function twct AS} should be replaced by objective function \eqref{eq: objective function cmax AS}, which minimizes the maximum end time of the jobs.
\begin{equation}
    \text{minimize~~} \max_{j \in \Jobs, m \in \Machines} ~ p_\jm \cdot (s_\jm + \ptime_j) \label{eq: objective function cmax AS}
\end{equation}
}

\begin{figure}[!t]
    \centering
    \includegraphics[width=\linewidth]{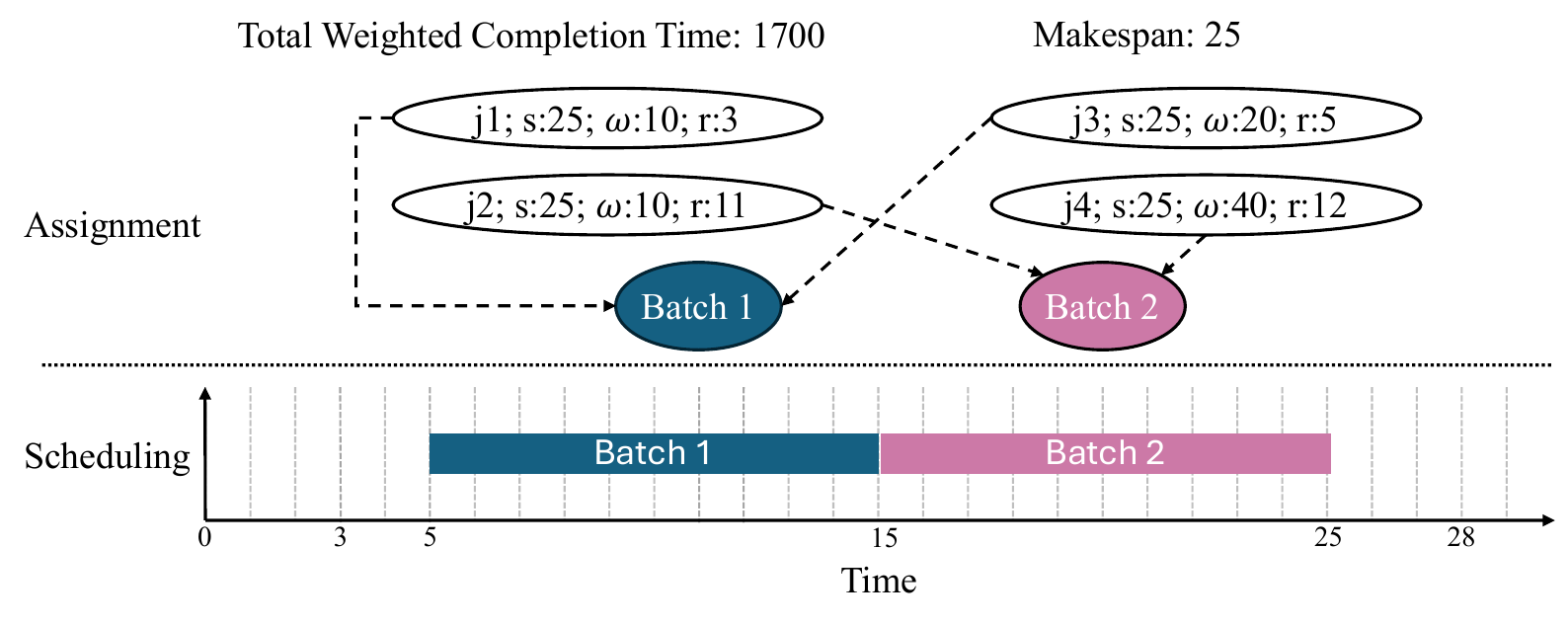}
    \caption{Optimal solution of the Assign-and-Schedule model}
    \label{fig: AS solution}
\end{figure}

Figure \ref{fig: AS solution} presents the optimal solution of the AS model minimizing the TWCT. Note that the structure of the solution changes, and there is a clear division between assignment and scheduling decisions.

%%%%%%%%%%%%%%%%%%%%%%%%%%%%%%%%%%%%%%%%%%%%%%%%%%%%%%%%%%%%%%%%%%%%%%%%
% Synchronized model
%%%%%%%%%%%%%%%%%%%%%%%%%%%%%%%%%%%%%%%%%%%%%%%%%%%%%%%%%%%%%%%%%%%%%%%%
\subsection{The Synchronized Model} \label{sec: synchronized model}

Model \ref{model: S} presents the Synchronized (S) model \nt{using the syntax of the commercial solver CP Optimizer. The same model can be implemented in open-source solvers with minimal changes}. It is based on the CP model by \citet{Cakici2024} for AM processes, which deals with a single compatible family; this paper extends their formulation to handle multiple incompatible families. Equation (\ref{eq: def interval job on machine in batch S}) defines an optional interval variable $x_\jm^b$, of size $\ptime_j$, that represents the option of job $j$ being batched in batch $b \in \Batches_{f_j}$ and processed on machine $m \in \Machines$. Equation (\ref{eq: def interval batch on machine S}) defines an optional interval variable $y_\bm$ of size $\ptime^{f^b}$ that represents the option of batch $b \in \Batches$ being processed on machine $m$. Equation (\ref{eq: def sequence var S}) defines an interval sequence variable $\varphi_m$ that represents the sequence of batches scheduled on machine $m$. Equation (\ref{eq: cumulative by family S}) defines a cumulative function $\cumulLoad_m^b$ that represents the load on machine $m$ due to the jobs in batch $b$.

Objective function (\ref{eq: objective function twct S}) minimizes the TWCT. Constraints (\ref{eq: job on one machine in one batch S}) assign each job to exactly one batch on exactly one machine. Constraints (\ref{eq: max one machine per batch S}) allow each batch to be processed in at most one machine. Constraints (\ref{eq: deprecated synchronization}) use the $\Synchronize\rt{(v, B)}$ global constraint, which \rt{receives} an interval variable $v$ and a set of interval variables $B$. If $v$ is present, this constraint attempts to synchronize it with any intervals in $B$ that are also present \cite{ibm2023cpoptimizer}, i.e., $\PresenceOf(v) = 1 \implies \StartOf(v) = \StartOf(b) \wedge \EndOf(v) = \EndOf(b), \forall ~ b \in B$ s.t. $\PresenceOf(b) = 1$. In practice, when the interval variable representing the job being assigned to a batch on a machine (i.e., $x_\jm^b$) is part of the solution, the $\Synchronize$ constraint attempts to align its start and end times with those of the batch on the machine (i.e., $y_\bm$), if it is also present. However, the $\Synchronize$ constraint only ensures synchronization if both interval variables are present; it does not inherently enforce the presence of $y_\bm$ when $x_\jm^b$ is scheduled. This subtle issue, where the presence of $x_\jm^b$ should logically imply the presence of $y_\bm$ is \nt{addressed in the original model by \citet{Cakici2024} by means of an additional constraints that defines the correct duration of the batches depending on the jobs assigned to it, if any. This constraint is necessary in their AM application as the duration of the batch depends on the size of the parts 3D-printed in it. However, this constraint is not required in the context of the diffusion area studied in this paper since the duration of the batches is already predefined. Nonetheless, the presence of the correct batches is still needed to be enforced. To resolve this, constraints (\ref{eq: fixed implication}) introduce an explicit implication constraint which ensures that, whenever $x_\jm^b$ is included in the solution, the corresponding batch interval $y_\bm$ is also present}. Constraints (\ref{eq: no overlapping S}) use the $\NoOverlap$ global constraint to ensure that the present interval variables in the machine sequence do not overlap, guaranteeing at most one batch being processed at a time on each machine. Finally, constraints (\ref{eq: always in by batch S}) guarantee that the load \nt{of the batches respect the maximum capacity.}

\begin{model}[!t]
\caption{The Synchronized Model Inspired by \cite{Cakici2024}} \label{model: S}
\renewcommand{\arraystretch}{1.3}
\begin{tabularx}{\linewidth}{@{}R@{}L@{}K@{}R@{}}
    \multicolumn{4}{@{}l}{Variables:} \\
    x_\jm^b & \multicolumn{3}{@{}L}{ \in \{[a,a + \ptime_j: \releaseTime_j \leq a \in \Horizon)\} \cup \{\perp\}, } \\
     & \multicolumn{2}{R@{}}{\forall ~ j \in \Jobs, m \in \Machines, b \in \Batches_{f_j}; }& \AddLabel{eq: def interval job on machine in batch S} \\
    y_\bm & \in \{[a,a + \ptime^{f^b}): a \in \Horizon\} \cup \{\perp\}, \quad \quad  & \forall ~ b \in \Batches, m \in \Machines; & \AddLabel{eq: def interval batch on machine S}\\
    \varphi_m & \in \Permutation\pp{\set{y_\bm}_{b \in \Batches}}, & \forall ~m \in \Machines. & \AddLabel{eq: def sequence var S}\\
    \multicolumn{4}{@{}l}{Functions:} \\
    \cumulLoad_m^b &= \sum_{j \in \Jobs_{f^b}} \pulse \pp{x^b_\jm, \size_j}, & \forall ~ m \in \Machines, b \in \Batches. & \AddLabel{eq: cumulative by family S}\\
\end{tabularx}
\noindent
\begin{tabularx}{\linewidth}{@{}XR@{}LR@{}}
\multicolumn{4}{@{}l}{Formulation:} \\
% Row 1: Objective function
    \multicolumn{3}{C}{\text{minimize~} \sum_{j \in \Jobs} \weight_j \cdot \sum_{b \in \Batches_{f_j}}\sum_{m \in \Machines} \EndOf\pp{x^b_\jm},} & \AddLabel{eq: objective function twct S} \\
% Subject to
    \multicolumn{4}{l}{subject to,} \\
% Each job in max one batch on max one machine
    \multicolumn{3}{L}{~~\forall ~j \in \Jobs:} & \multirow{2}{*}{\AddLabel{eq: job on one machine in one batch S}} \\
    \multicolumn{3}{L@{}}{\phantom{\sum_{m \in \Machines}}\sum_{b \in \Batches_{f_j}}\sum_{m \in \Machines} \PresenceOf \pp{x_\jm^b} = 1;}    & \\
% Each bacth on max one machine
    \multicolumn{3}{L}{~~\forall ~b \in \Batches:} & \multirow{2}{*}{\AddLabel{eq: max one machine per batch S}} \\
    \multicolumn{3}{L@{}}{\phantom{\sum_{m \in \Machines}\sum_{b \in \Batches_{f_j}}}\sum_{m \in \Machines} \PresenceOf \pp{y_\bm} \leq 1;}  & \\
% Synchronization
    \multicolumn{3}{L}{~~\forall ~j \in \Jobs, b \in \Batches_{f_j}, m \in \Machines} & \\
    \multicolumn{2}{E@{}}{\Synchronize} & \pp{x^b_\jm, \{y_\bm\}};  &  \AddLabel{eq: deprecated synchronization}\\
    \multicolumn{2}{E@{}}{\PresenceOf} & \pp{x^b_\jm} \Rightarrow \PresenceOf \pp{y_\bm};  & \AddLabel{eq: fixed implication} \\
% No overlap of batches
    \multicolumn{3}{L}{~~\forall ~ m \in \Machines: } & \multirow{2}{*}{\AddLabel{eq: no overlapping S}} \\
    \multicolumn{2}{R@{}}{\NoOverlap} & \pp{\varphi_m}; & \\
% AlwaysIn by batch
    % \multicolumn{3}{L}{~~ \forall ~ j \in \Jobs, m \in \Machines, b \in \Batches_{f_j}:} & \multirow{2}{*}{\AddLabel{eq: always in by batch S}} \\
    % \multicolumn{2}{R@{}}{\AlwaysIn} & \bigl(\cumulLoad_m^b, x_\jm^b, l_{f^b}, u_{f^b}\bigr).  & \\
    \multicolumn{3}{L}{\forall ~ m \in \Machines, b \in \Batches: \quad \cumulLoad_m^b \leq u_f^b.} & \AddLabel{eq: always in by batch S} \\
\end{tabularx}
\end{model}

\nt{To minimize the makespan, objective function \eqref{eq: objective function twct S} should be replaced by objective function \eqref{eq: objective function cmax S}, which minimizes the maximum end time of the jobs.
\begin{equation}
    \text{minimize~~} \max_{b \in \Batches, m \in \Machines} ~ \EndOf(y_\bm) \label{eq: objective function cmax S}
\end{equation}
}

\begin{figure}[!t]
    \centering
    \includegraphics[width=\linewidth]{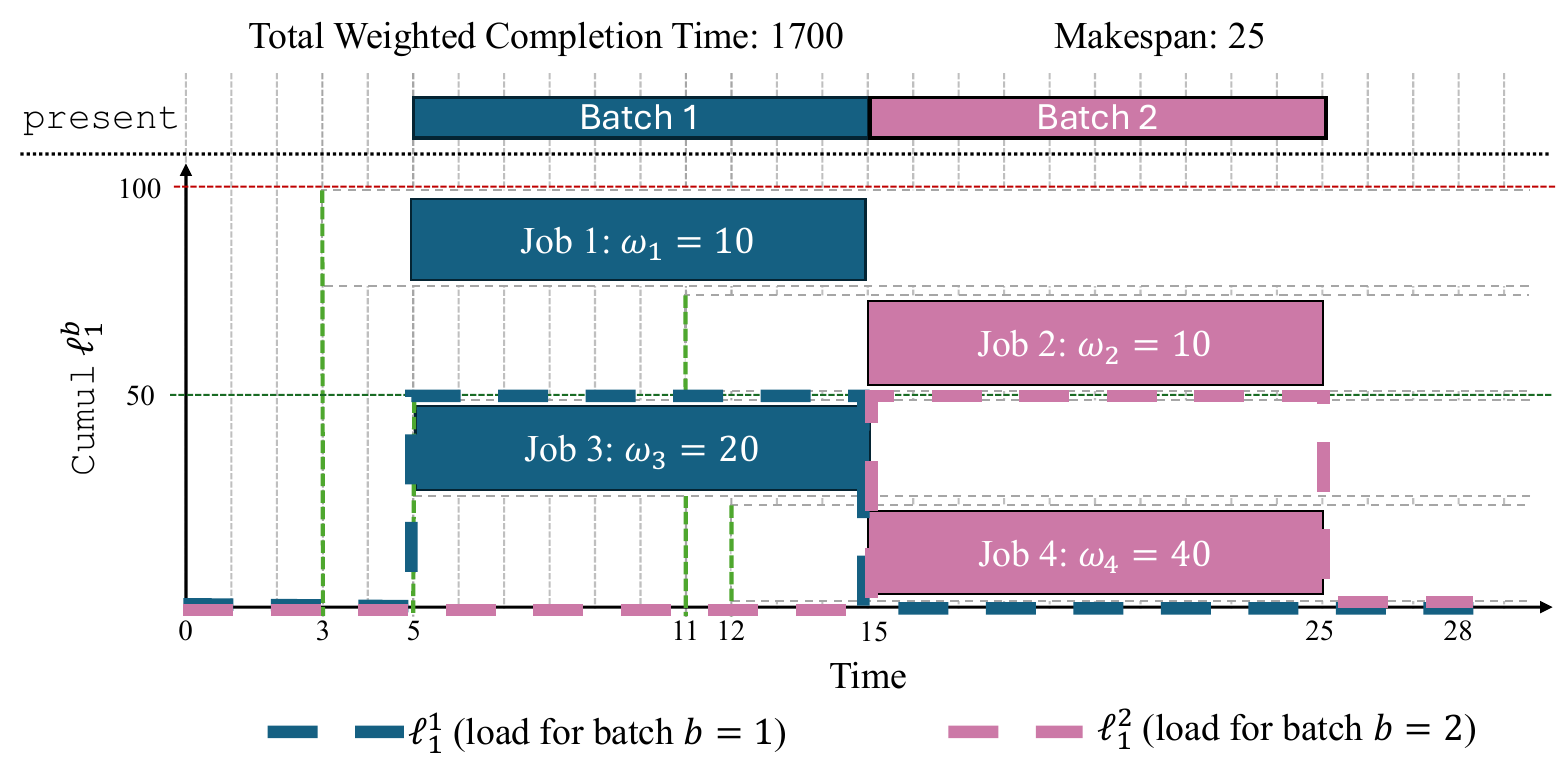}
    \caption{Optimal solution of the Synchronized model}
    \label{fig: S solution}
\end{figure}

Figure \ref{fig: S solution} shows the optimal solution of the S model. \nt{Note how the jobs belong now to batches and the cumulative loads are now tallied by batches.}

% \nt{Synchronization constraints \eqref{eq: deprecated synchronization} are only available in CP Optimizer. However, the implementation in other solvers is straight forward via implication constraints: $\PresenceOf(x^b_\jm) \Rightarrow \StartOf(v) = \StartOf(x^b_\jm)$.}

%%%%%%%%%%%%%%%%%%%%%%%%%%%%%%%%%%%%%%%%%%%%%%%%%%%%%%%%%%%%%%%%%%%%%%%%
% Redundant Synchronized model
%%%%%%%%%%%%%%%%%%%%%%%%%%%%%%%%%%%%%%%%%%%%%%%%%%%%%%%%%%%%%%%%%%%%%%%%

\subsection{The Redundant Synchronized Model} \label{sec: redundant synchronized}

Model \ref{model: RS} presents the Redundant Synchronized (RS) model \nt{using the syntax of the commercial solver CP Optimizer. The same model can be implemented in open-source solvers with minimal changes}.

\begin{model}[!t]
\caption{Redundant Synchronized model} \label{model: RS}
\renewcommand{\arraystretch}{1.3}
\begin{tabularx}{\linewidth}{@{}R@{}L@{}K@{}R@{}}
% Variables
    \multicolumn{4}{@{}l}{Variables:} \\
    x_j & \in \{[a,a + \ptime_j: \releaseTime_j \leq a \in \Horizon)\}, & \forall ~j \in \Jobs; & \AddLabel{eq: def interval job RS} \\
    x_\jm & \in \{[a,a + \ptime_j: a \in \Horizon) \} \cup \{\perp\}, & \forall ~ j \in \Jobs, m \in \Machines; & \AddLabel{eq: def interval job on machine RS}\\
    x_\jm^b & \in \{[a,a + \ptime_j: a \in \Horizon)\} \cup \{\perp\}, & \\
     & \multicolumn{2}{@{}R}{\forall ~ j \in \Jobs, m \in \Machines, b \in \Batches_{f_j};} & \AddLabel{eq: def interval job on machine in batch RS}\\
    y_b & \in \{[a,a + \ptime^{f^b}: a \in \Horizon)\} \cup \{\perp\}, \quad~ & \forall ~b \in \Batches; & \AddLabel{eq: def interval batch RS} \\
    y_\bm & \in \{[a,a + \ptime^{f^b}: a \in \Horizon)\} \cup \{\perp\}, & \forall ~ b \in \Batches, m \in \Machines; & \AddLabel{eq: def interval batch on machine RS}\\
    \varphi_m & \in \Permutation\pp{\set{y_\bm}_{b \in \Batches}}, & \forall ~m \in \Machines. & \AddLabel{eq: def sequence var RS}\\
% Sets of variables
    \multicolumn{4}{@{}l}{Sets of variables to synchronize with $x_\jm^b$:} \\
    V_\jm^b & \multicolumn{2}{@{}L}{= \{y_b, y_\bm, x_\jm \}, \quad \quad \forall ~ j \in \Jobs, m \in \Machines, b \in \Batches_{f_j};} & \AddLabel{eq: def set V}\\
% Functions
    \multicolumn{4}{@{}l}{Functions:} \\
    \machineStateFamily_m &: \StateFunction & \forall ~m \in \Machines; & \AddLabel{eq: state by family RS}\\
    \cumulLoad_{mf} &= \sum_{j \in \Jobs_f} \pulse \pp{x_\jm, \size_j}, & \forall ~ m \in \Machines, f \in \Families. & \AddLabel{eq: cumulative by family RS}\\
\end{tabularx}
% Formulation
\begin{tabularx}{\linewidth}{@{}XR@{~}LR@{}}
\multicolumn{4}{@{}l}{Formulation:} \\
% Row 1: Objective function
    \multicolumn{3}{C}{\text{minimize ~ } \sum_{j \in \Jobs} \weight_j \cdot \EndOf\pp{x_j},} & \AddLabel{eq: objective function twct RS} \\
% Subject to
\multicolumn{4}{l}{subject to,} \\
% One job on one machine
    \multicolumn{3}{L}{~~\forall ~j \in \Jobs:} & \multirow{2}{*}{\AddLabel{eq: job on one machine RS}} \\
    \multicolumn{2}{R@{}}{\Alternative} & \pp{x_j, \set{x_\jm}_{m \in \Machines}};  & \\
% If job on machine => job on machine in a batch
    \multicolumn{3}{L}{~~\forall ~ j \in \Jobs, m \in \Machines:} & \multirow{2}{*}{\AddLabel{eq: job on machine in one batch RS}} \\
    \multicolumn{2}{R@{}}{\Alternative} & \Bigl(x_\jm, \set{x^b_\jm}_{b \in \Batches_{f_j}}\Bigr);  & \\
% If batch => batch on one machine
    \multicolumn{3}{L}{~~\forall ~b \in \Batches: } & \multirow{2}{*}{\AddLabel{eq: each batch on one machine RS}} \\
    \multicolumn{2}{R@{}}{\Alternative} & \pp{y_b , \set{y_\bm}_{m \in \Machines}}; & \\
% Synchronization
    \multicolumn{3}{L}{~~\forall ~j \in \Jobs, b \in \Batches_{f_j}, m \in \Machines, v \in V_\jm^b:} & \\
    \multicolumn{2}{E@{}}{\PresenceOf} & \pp{x^b_\jm} \Rightarrow \PresenceOf\pp{v};  & \AddLabel{eq: if then} \\
    % \multicolumn{2}{R@{}}{\StartAtStart} & \bigl(x^b_\jm, v\bigr);  & \AddLabel{eq: start at start} \\
    % \multicolumn{2}{R@{}}{\EndAtEnd} & \bigl(x^b_\jm, v\bigr) ;  & \AddLabel{eq: end at end} \\
% Synchronization
    \multicolumn{3}{L}{~~\forall ~j \in \Jobs, b \in \Batches_{f_j}, m \in \Machines:} & \\
    \multicolumn{2}{E@{}}{\Synchronize} & \pp{x^b_\jm, V^b_\jm};  & \AddLabel{eq: synchronize RS} \\
% No overlap of batches
    \multicolumn{3}{L}{~~\forall ~ m \in \Machines: } & \multirow{2}{*}{\AddLabel{eq: no overlapping RS}} \\
    \multicolumn{2}{R@{}}{\NoOverlap} & \pp{\varphi_m}; & \\
% AlwaysEqual by family
    \multicolumn{3}{L}{~~ \forall ~j \in \Jobs, m \in \Machines:} & \multirow{2}{*}{\AddLabel{eq: always equal by family RS}} \\
    \multicolumn{2}{R@{}}{\AlwaysEqual} & \bigl(\machineStateFamily_m, x_\jm, f_j \bigr);  & \\
% AlwaysIn by batch
    % \multicolumn{3}{L}{~~ \forall ~ m \in \Machines, f \in \Families, j \in \Jobs_{f_j}:} & \multirow{2}{*}{\AddLabel{eq: always in by family RS}} \\
    % \multicolumn{2}{R@{}}{\AlwaysIn} & \bigl(\cumulLoad_{mf}, x_\jm, l_f, u_f\bigr).  & \\
    \multicolumn{3}{L}{\forall ~ m \in \Machines, f \in \Families: \quad \cumulLoad_{mf} \leq u_f.} & \AddLabel{eq: cumulative bound by family RS} \\
\end{tabularx}
\end{model}

Equation (\ref{eq: def interval job RS}) defines an interval variable $x_j$ of size $\ptime_j$ that represents job $j \in \Jobs$. Equation (\ref{eq: def interval job on machine RS}) defines an optional interval variable $x_\jm$ that represents the option of processing job $j$ on machine $m \in \Machines$. Equation (\ref{eq: def interval job on machine in batch RS}) defines an optional interval variable $x_\jm^b$, also of size $\ptime_j$, that represents the option of job $j$ being batched in batch $b \in \Batches_{f_j}$ and processed on machine $m$. Equation (\ref{eq: def interval batch RS}) defines an optional interval variable $y_b$ of size $\ptime^{f^b}$ that represents batch $b \in \Batches$. Equation (\ref{eq: def interval batch on machine RS}) defines an optional interval variable $y_\bm$ of size $\ptime^{f^b}$ that represents the option of batch $b$ being processed on machine $m$. Equation (\ref{eq: def sequence var RS}) defines an interval sequence variable $\varphi_m$ that represents the sequence of batches scheduled on machine $m$. Equation (\ref{eq: def set V}) defines the set $V_\jm^b$ of interval variables that have to be synchronized with interval $x_\jm^b$, if it is present. Equation (\ref{eq: state by family RS}) defines the unique state $\machineStateFamily_m$ of the machine $m$ at every point in time. Equation (\ref{eq: cumulative by family RS}) defines a cumulative function $\cumulLoad_{mf}$ that represents the load on machine $m$ with jobs of family $f$.

Objective function (\ref{eq: objective function twct RS}) minimizes the TWCT. Constraints (\ref{eq: job on one machine RS}) ensure that each job is processed on exactly one machine. This can be seen as an assignment of jobs to machines, but now using the global $\Alternative$ constraint, which exploits the structure of the problem to help the search process. Similarly, constraints (\ref{eq: job on machine in one batch RS}) ensure that if a job is processed on a given machine, it is processed on such machine in exactly one batch. This can be seen as another assignment of jobs to batches on specific machines, but now using CP scheduling logic. Constraints (\ref{eq: each batch on one machine RS}) guarantee that each batch is processed on exactly one machine. This can be seen as another assignment of batches to machines with global constraints. The previous three constraint groups are completely independent and allow the machine assigned to each job, the batch where it is assigned, and the machine assigned to the batch, to be completely unrelated. Constraints (\ref{eq: if then}) remedy this and link them all together, ensuring that if the interval variable $x^b_\jm$ for the job $j \in \Jobs$ in the batch $b \in \Batches_{f_j}$ on the machine $m \in \Machines$ is present in the final solution, then it must be the case that all the intervals in $V^b_\jm$ must be present in the solution as well (i.e., the batch interval $y_b$, the interval of the batch on the machine $y_\bm$, and the interval of the job on the machine $x_\jm$). Furthermore, constraints (\ref{eq: synchronize RS}) synchronize the start and end of the interval $x^b_\jm$ with the intervals in $V^b_\jm$, if they are present in the final solution.
Constraints (\ref{eq: no overlapping RS}) ensure that only one batch is processed at a time on each machine.
Constraints (\ref{eq: always equal by family RS}) ensure that jobs of only one family are processed at any time on the machines. \nt{This global constraint is only available in CP Optimizer. When implementing in other solvers, removing this constraint does not impact the feasibility of the solution achieved since constraints \eqref{eq: no overlapping RS} already ensure this as well. }Finally, constraints (\ref{eq: cumulative bound by family RS}) guarantee that the load on the machines is within the required batch size limits of the family being processed.

\nt{To minimize the makespan, objective function \eqref{eq: objective function twct RS} should be replaced by objective function \eqref{eq: objective function cmax NS}, which minimizes the maximum end time of the jobs.}

\subsection{Breaking symmetries} \label{sec: symmetry-breaking}

Symmetries in the search space could delay the search process to prove optimality. Hence, breaking these symmetries could potentially benefit the search process. Since the \textit{number} of the batch is irrelevant for the solution, the following symmetry-breaking constraints, borrowed from \citet{Nascimento2023}, can be included in the AS and RS models.
\begin{figure}[!t]
\noindent
\begin{tabularx}{\linewidth}{@{}XR@{}LR@{}}
    \multicolumn{3}{L}{~~ \forall ~ f \in \Families, b \in \Batches_f \setminus \set{\min \Batches_f}:} & \\
    & \PresenceOf & \pp{y_b} \leq \PresenceOf\pp{y_{b-1}};  & \AddLabel{eq: symmetry-breaking presence}\\
    \multicolumn{3}{C}{\StartBeforeStart \pp{y_{b-1}, y_{b}}.}  & \AddLabel{eq: symmetry-breaking time} 
% \bottomrule
\end{tabularx}
\end{figure}
Constraints (\ref{eq: symmetry-breaking presence}) ensure that the batches of a family are used incrementally, i.e., if a batch $b$ is used, is because all the previous batches have already been used. Constraints (\ref{eq: symmetry-breaking time}) go further by ensuring that the start times of the batches also increase with the batches. They use the global constraint $\StartBeforeStart$, which ensures that the start time of the first interval variable provided is before the start time of the second interval provided. \nt{This constraint can be implemented in other solvers using reification.}

\section{Computational experiments and results} \label{sec: experiments and results}

% \begin{table}[!t]
%     \caption{The Structure of the instances.}
%     \label{tab: instances structure}
%     \centering
%     \begin{tabular}{lcl}
%         \toprule
%             \textbf{Category} & \textbf{Levels} & \textbf{Definitions}\\
%         \midrule
%             Jobs                & 4 & $|\Jobs| \in \set{50, 100, 150, or 200}$ \\
%             Families            & 5 & $|\Families| \in \set{4, 5, 6, 8, 10}$ \\
%             Machines            & 5 & $|\Machines| \in \set{4, 5, 6, 8, 10}$ \\
%             Processing times    & 1 & $\ptime^f \sim U\set{[10]}$ \\
%             Job sizes           & 1 & $\size_j \sim U\set{[25]}$ \\
%             Job weights         & 1 & $\weight_j \sim U\set{[10]}$ \\
%             Release times       & 1 & $\releaseTime_j \sim U\set{[C_{max, lb}]}$\\
%             \midrule
%             \# of instance classes & 100 & \\
%             \# of instances per class & 10 & \\
%             \cmidrule{1-2}
%             Total \# of instances & 1,000 \\
%         \bottomrule
%     \end{tabular}
% \end{table}

\begin{figure*}[htbp]
    \centering
    \begin{subfigure}[b]{0.475\linewidth}
        \centering
        \includegraphics[width=\textwidth]{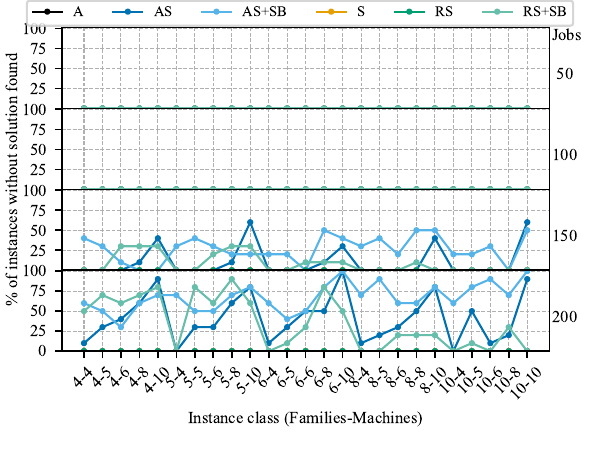}
        \vspace{-2em}
        \caption{\% of instances without solution}
        \label{fig: CP Opt TWCT no solution}
    \end{subfigure}
    \quad
    \begin{subfigure}[b]{0.475\linewidth}
        \centering
        \includegraphics[width=\textwidth]{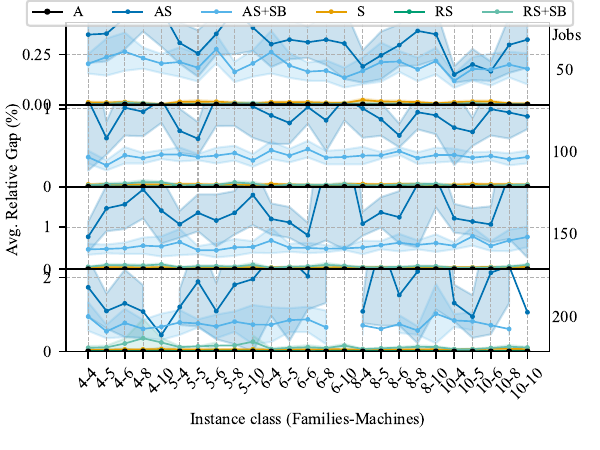}
        \vspace{-2em}
        \caption{Average \rt{relative} gap (\%) with 95\% CI}
        \label{fig: CP Opt TWCT Gap}
    \end{subfigure}
    \caption{Results minimizing the TWCT with CP Optimizer}
    \label{fig: CP Opt TWCT}
\end{figure*}

\begin{figure*}[htbp]
    \centering
    \begin{subfigure}[b]{0.475\linewidth} % Adjust width to fit three subfigures
        \centering
        \includegraphics[width=\textwidth]{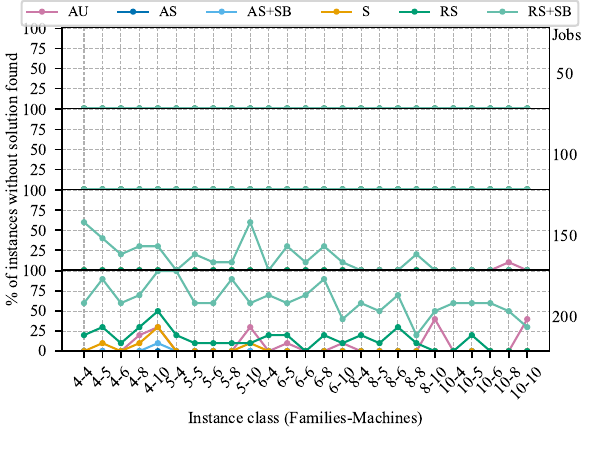}
        \vspace{-2em}
        \caption{\% of instances without solution}
        \label{fig: CP-SAT TWCT no solution}
    \end{subfigure}
    \quad
    \begin{subfigure}[b]{0.475\linewidth} % Adjust width to fit three subfigures
        \centering
        \includegraphics[width=\textwidth]{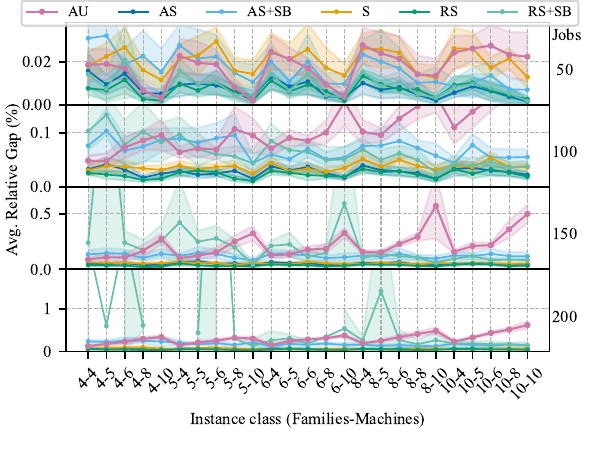}
        \vspace{-2em}
        \caption{Average \rt{relative} gap (\%) with 95\% CI}
        \label{fig: CP-SAT TWCT Gap}
    \end{subfigure}
    \caption{Results minimizing the TWCT with CP-SAT}
    \label{fig: CP-SAT TWCT}
\end{figure*}

\begin{figure*}[htbp]
    \centering
    \begin{subfigure}[b]{0.475\linewidth}
        \centering
        \includegraphics[width=\textwidth]{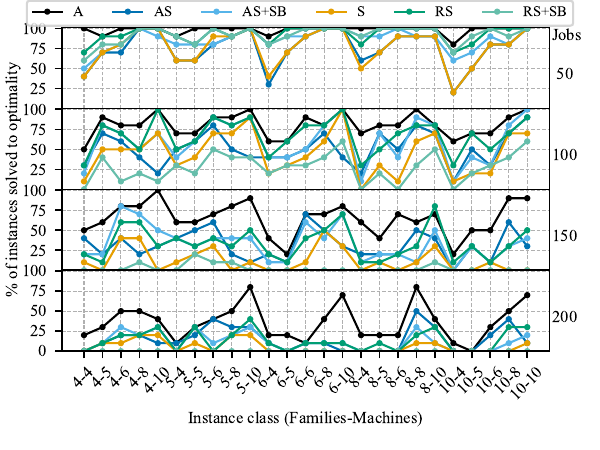}
        \vspace{-2em}
        \caption{\% of instances solved to optimality}
        \label{fig: CP Opt Cmax optimality}
    \end{subfigure}
    \quad
    \begin{subfigure}[b]{0.475\linewidth} % Adjust width to fit three subfigures
        \centering
        \includegraphics[width=\textwidth]{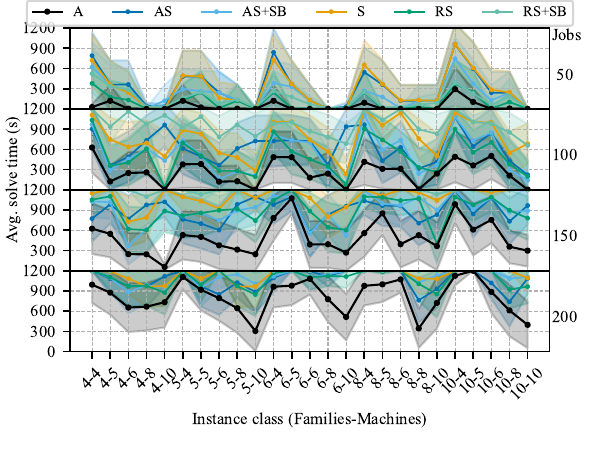}
        \vspace{-2em}
        \caption{Average solve time (s)}
        \label{fig: CP Opt Cmax Time}
    \end{subfigure}
    \\\vspace{1em}
    \begin{subfigure}[b]{0.475\linewidth} % Adjust width to fit three subfigures
        \centering
        \includegraphics[width=\textwidth]{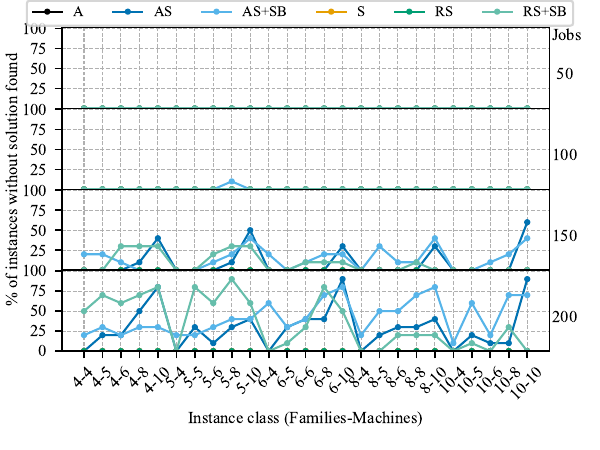}
        \vspace{-2em}
        \caption{\% of instances without solution}
        \label{fig: CP Opt Cmax no solution}
    \end{subfigure}
    \quad
    \begin{subfigure}[b]{0.475\linewidth}
        \centering
        \includegraphics[width=\textwidth]{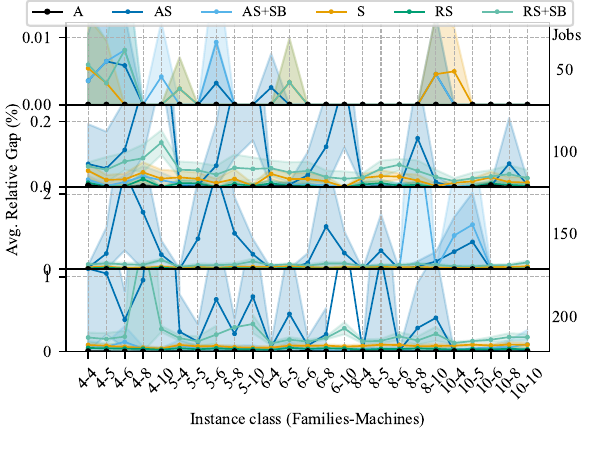}
        \vspace{-2em}
        \caption{Average \rt{relative} gap (\%) with 95\% CI}
        \label{fig: CP Opt Cmax Gap}
    \end{subfigure}
    \caption{Results minimizing the $\makespan$ with CP Optimizer.}
    \label{fig: CP Opt Cmax}
\end{figure*}

\begin{figure*}[htbp]
    \centering
    \begin{subfigure}[b]{0.475\linewidth}
        \centering
        \includegraphics[width=\textwidth]{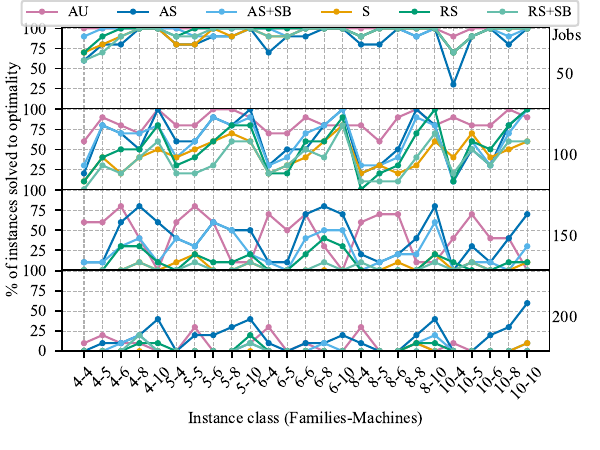}
        \vspace{-2em}
        \caption{\% of instances solved to optimality}
        \label{fig: CP-SAT Cmax optimality}
    \end{subfigure}
    \quad
    \begin{subfigure}[b]{0.475\linewidth} % Adjust width to fit three subfigures
        \centering
        \includegraphics[width=\textwidth]{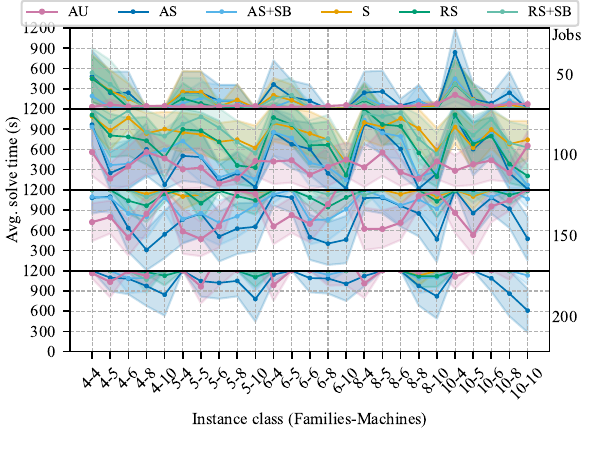}
        \vspace{-2em}
        \caption{Average solve time (s)}
        \label{fig: CP-SAT Cmax Time}
    \end{subfigure}
    \\\vspace{1em}
    \begin{subfigure}[b]{0.475\linewidth} % Adjust width to fit three subfigures
        \centering
        \includegraphics[width=\textwidth]{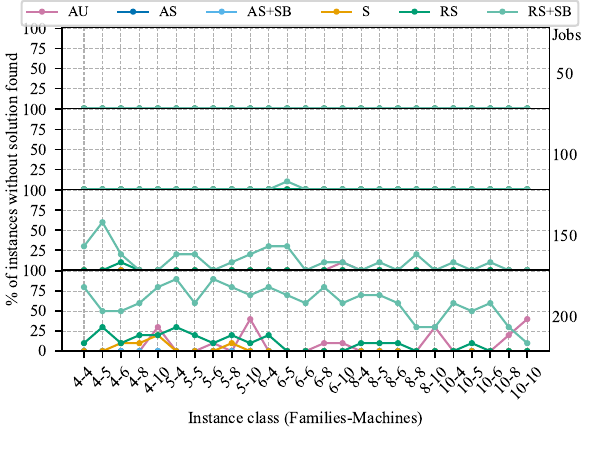}
        \vspace{-2em}
        \caption{\% of instances without solution}
        \label{fig: CP-SAT Cmax no solution}
    \end{subfigure}
    \quad
    \begin{subfigure}[b]{0.475\linewidth}
        \centering
        \includegraphics[width=\textwidth]{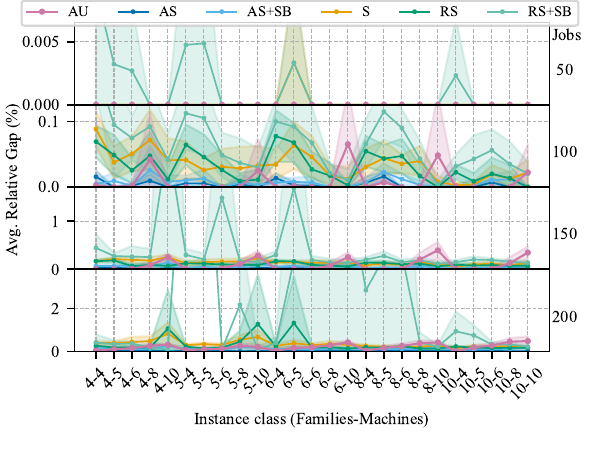}
        \vspace{-2em}
        \caption{Average \rt{relative} gap (\%) with 95\% CI}
        \label{fig: CP-SAT Cmax Gap}
    \end{subfigure}
    \caption{Results minimizing the $\makespan$ with CP-SAT.}
    \label{fig: CP-SAT Cmax}
\end{figure*}
\nt{
The computational experiments consider \nt{1,000} instances generated following a similar process to that of \citet{Cakici2013} and \citet{Ham2017-CP-p-batch-incompatible}. The number of jobs in each instance is one of four possible values: \nt{50, 100, 150, or 200}. \nt{Likewise, the number of families and the number of machines in each instance can be one of five possible values: 4, 5, 6, 8, or 10}. Overall, there are $\nt{100} = 4 \times 5 \times 5$ instance classes, and for each class, 10 random instances were generated, totaling \nt{1,000} instances.

For every instance, the processing time for each family $f \in \Families$ and the weight for each job $j \in \Jobs$ follow the distribution $ \ptime^f, \weight_j \sim U\set{[\nt{10}]}$, where $U\set{[\varsigma]}$ denotes the discrete uniform distribution over the set $[\varsigma]=\{1,2,\dots,\varsigma\}$. Job sizes are drawn from $\size_j \sim U\set{[25]}$. Release times are generated by first computing a lower bound of the overall makespan as 
\[
C_{\max, lb} = \frac{\sum_{f \in \Families} \ptime^f \cdot \ceil{N_f/u_f}}{|\Machines|},
\]
and then setting $ \releaseTime_j \sim U\set{[C_{\max, lb}]}$.
Finally, the maximum batch size for each family \rt{follows the distribution $u_f \sim U\{\Omega_f\}$,
where $\Omega_f = \{u \in \Omega : u \geq \max_{j \in \Jobs_f} \size_j\}$ and $\Omega = \{10, 15, 20, 25, 30, 35, 40, 45, 50, 75, 100, 125, 150 \}$. This approach ensures that instances are feasible by construction, allowing maximum batch sizes for each family that are larger than the size of any of its jobs.}

\subsection*{Models and Solvers}

Each instance is solved by minimizing two objectives independently—TWCT and $\makespan$—using the following seven models:
\begin{itemize}
    \item A: Aligned model by \citet{Ham2017-CP-p-batch-incompatible}.
    \item AU: Automaton model from Section \ref{sec: automaton model}.
    \item AS: Assign-and-Schedule model from Section \ref{sec: assign and schedule model}.
    \item AS+SB: AS model with symmetry-breaking (SB) constraints.
    \item S: Synchronized model from Section \ref{sec: synchronized model}.
    \item RS: Redundant Synchronized model from Section \ref{sec: redundant synchronized}.
    \item RS+SB: RS model with symmetry-breaking constraints.
\end{itemize}

To showcase the flexibility of our modeling approach, we used two solvers:
\begin{itemize}
    \item Commercial solver: IBM ILOG CP Optimizer \cite{Laborie2018} from \textsc{CPLEX} v.22.1.1,
    \item Open-source solver: Google OR-Tools CP-SAT \cite{ortools, cpsatlp} v.9.11.
\end{itemize}

Table \ref{tab: models per solver} summarizes the models implemented in each solver. The A model was only implemented in CP Optimizer because its state functions, $\AlwaysEqual$ global constraint, and alignment parameters are only available in it. On the other hand, the AU model was only implemented in CP-SAT because the $\Automaton$ global constraint required is not available in CP Optimizer (although it can be implemented in Choco-solver \cite{ChocoSolver} and MiniZinc \cite{MiniZinc-Nethercote2007}). The AS, S, and RS models (and their variants with SB constraints) are implemented in both solvers.
\begin{table}[!t]
    \caption{Models implemented by solver}
    \label{tab: models per solver}
    \centering
    \begin{tabular}{llr}
        \toprule
            \textbf{Solver} & \textbf{Models} & \textbf{Total}\\
        \midrule
            CP Optimizer & ~A~, AS, AS+SB, S, RS, RS+SB & 6\\
            CP-SAT & AU, AS, AS+SB, S, RS, RS+SB & 6\\
        \bottomrule
    \end{tabular}
\end{table}

Overall, a total of 24,000 experiments were conducted: $1,000\ \text{instances} \times 2\ \text{objectives} \times 6\ \text{models per solver} \times 2\ \text{solvers}$. All the models were implemented in \nt{\textsc{Python} 3.10} using the appropriate \textsc{Python} interfaces \cite{ibm2023cpoptimizerpython}. All the experiments were run on the PACE Phoenix cluster \cite{PACE} using machines with Red Hat Enterprise Linux Server release 7.9 (Maipo), dual Intel\textregistered~Xeon\textregistered~ Gold 6226 CPU @ 2.70GHz (24 cores), \nt{72} GB RAM, and parallelization of up to three experiments concurrently (each experiment using 8 cores and \nt{24} GB RAM). Each experiment was allocated a time limit of \nt{20} minutes (\nt{1,200} seconds) to match the practical constraints of a diffusion scheduling system in the semiconductor industry \cite{Ham2017-CP-p-batch-incompatible}.

Figures \ref{fig: CP Opt TWCT}--\ref{fig: CP-SAT Cmax} summarize our results. Figures \ref{fig: CP Opt TWCT} and \ref{fig: CP-SAT TWCT} focus on the results for the TWCT, while Figures \ref{fig: CP Opt Cmax} and \ref{fig: CP-SAT Cmax} focus on the $\makespan$ objective. Each graph presents the following characteristics:
\begin{itemize}
    \item Instance Grouping: the stacked line charts group instances by the number of jobs (indicated on the right-hand side) and by the combination of families and machines along the horizontal axis.
    \item Data Aggregation: each marker represents the aggregated result (over 10 instances) of a given instance class.
    \item Model Comparison: each series corresponds to one of the six models implemented in each solver, yielding metrics from 6,000 experiments per graph.
\end{itemize}

The metrics reported for each model include:
\begin{enumerate}
    \item Percentage of instances solved to optimality.
    \item Percentage of instances where no solution was found
    \item Average \rt{relative} gap (\%) with its 95\% confidence interval (CI).
    \item Average solve time (s) with its 95\% CI.
\end{enumerate}

The \rt{relative} gap for each instance is computed as follows. First, the minimum objective value $z_{i,obj}^*$ for each instance $i$ and objective $obj \in \{TWCT, \makespan\}$ \rt{is computed from the ones  obtained by all the models in all the solvers}. Then, for a given model and solver, the \rt{relative} gap of the instance is defined as $\frac{|z_{i,obj,\text{model, solver}} - z_{i,obj}^*|}{|z_{i,obj}^*|}$. \rt{It is relative because it is with respect to the \textit{best known} objective function value instead of the \textit{optimal} value.}

}

\nt{
\subsection{Results with the TWCT objective}

Figures \ref{fig: CP Opt TWCT} and \ref{fig: CP-SAT TWCT} present the results minimizing the TWCT with CP Optimizer and CP-SAT, respectively. Under the TWCT objective, all the models reached the time limit of 1,200 seconds. Hence, the graphs that show the average solve times of the models for both solvers are omitted. Additionally, no model declared optimality on any instance, with the exception of the AS+SB model implemented in CP-SAT, which declared optimality in only one instance with 50 jobs, 4 families, and 10 machines. For this reason, the graphs that show the percentage of instances solved to optimality for both solvers are also omitted.

Figure \ref{fig: CP Opt TWCT} focus\rt{es} on the results obtained with CP Optimizer. In particular, Figure \ref{fig: CP Opt TWCT no solution} reports the percentage of instances for which each model failed to find a solution. For instances with 50 and 100 jobs, all models successfully found solutions. However, as the number of jobs increases, the failure rate rises notably for the AS, AS+SB, and RS+SB models. The AS model (shown in dark blue) exhibits a distinct seasonal pattern: its performance deteriorates as the number of machines increases, peaking at 10 machines—for instance, no solutions were found for any instance with 200 job\rt{s}, 6 families, and 10 machines. The AS+SB model shows a similar seasonal behavior around 10 machines but also fails to find solutions in instances with fewer machines. In contrast, the RS+SB model follows this seasonal behavior yet performs better in larger instances, likely due to randomness in the instance generation. Notably, the A, S, and RS models were able to find solutions in all instances.

Figure \ref{fig: CP Opt TWCT Gap} presents the average \rt{relative} gap (\%) along with its 95\% confidence interval for each model in CP Optimizer. Although the A model did not prove optimality within the time limit in any instance, it consistently finds the best solutions, evident by the constant 0\% \rt{relative} gap of the black series. \rt{The RS and RS+SB models also obtained relative gaps close to 0\%, and this is why their series are presented under the black one. }In contrast, the AS and AS+SB models exhibit the worst average gaps, with wider confidence intervals. This is partly due to their failure to find solutions in several instances, resulting in fewer data points for averaging. Notably, a series break for these models appears at 200 jobs, 6 families, and 10 machines, reflecting a complete failure to solve any instance in that class. Furthermore, the consistently lower gaps of the AS+SB model compared to the AS model underscore the beneficial impact of incorporating SB constraints in the search process.

The AS and AS+SB models perform worse in CP Optimizer because of their structure.
}
Although the beauty of these models is their clear distinction between assignment decisions with binary variables and scheduling decisions with interval variables and global constraints, this separation reduces the ability of CP Optimizer to prune the domains of the scheduling variables. The $\NoOverlap$ constraints, in charge of handling the scheduling logic, provide the CP engine with a global view of the disjunctive constraints and effectively reduce the search space. However, constraints (\ref{eq: binary assignment})-(\ref{eq: capacity ub}), in charge of handling the assignment while respecting the batch limits, are local in nature, reduce the overall propagation, and provide weak information to the global scheduling constraints. 
\nt{As a result, this myopic structure heavily impacts these models' performance. Instead, the S and RS models, which replace the binary assignment with interval variables and global constraints, provide better information to the search process, allowing them to find solutions to all the instances and find solutions that are close to be the best. 

The extra SB constraints of the RS+SB impact this model's ability to find solutions in larger instances because they become harder to satisfy within the time limit. Note that the SB constraints impose additional requirements on the identifier of the batches. This is only useful for constraints \eqref{eq: each batch on one machine RS} to determine which batches to use. However, the implication constraints \eqref{eq: if then} only enforce the presence of a batch if the interval of any job in such batch on any machine is also present, not vice-versa. Hence, the SB constraints don't really provide better information to the CP engine during the search process to determine which job intervals to choose, precisely because of this directionality of the implications. This in turn makes the SB constraints harder to satisfy, causing more failures during the search.

Figure \ref{fig: CP-SAT TWCT} focus\rt{es} on the results of CP-SAT, which combines CP methods and satisfiability (SAT) methods in the search process. SAT methods benefit from binary structures in the problem. For this reason, in this solver, the AS model is able to find solutions on all the instances, as evidenced in Figure \ref{fig: CP-SAT TWCT no solution}. The distinction between assignment and scheduling decisions of the AS model provides the proper structure for both CP and SAT methods to properly work, allowing the AS model to find solutions in all the instances; and even declare optimality in one of them. In contrast, the RS+SB model is unable to find solutions in most of the largest instances, precisely because of the additional SB constraints that become harder to satisfy within the time limit. The RS model is unable to find solutions in the largest instances with 200 job\rt{s}, precisely because of the amount of job-in-batch decisions that need to be made. On the other hand, the AU model clearly shows seasonal behavior around 10 machines.

When examining the average \rt{relative} gaps in CP-SAT from Figure \ref{fig: CP-SAT TWCT Gap}, it becomes evident that no model consistently matches the best results obtained by the A model in CP Optimizer. Nonetheless, the gaps are very close. The AU gap increases with the size of the instances. This is because of the $\tau_{mt}$ variables that represent the transition values provided to the automaton at every time step of the scheduling horizon, which increases with the number of jobs. Additionally, the $\Element$ constraints set the appropriate transition values at the start times of the jobs on each machine. However, they do so only if the jobs are actually processed on such machine. While this could be seen as a reified constraint, it is not. Instead, this is indirectly achieved with the modified start times $\hat{s}_\jm$ and modified transition values $\hat{\tau}_\jm$. When the instances grow, more of these modified decisions have to be made, negatively impacting the performance of these $\Element$ constraints, and causing the increase on the average \rt{relative} gap.

In CP-SAT, it becomes evident that satisfying the SB constraints within the time limit is difficult for both the AS and RS models. The SB constraints not only impose additional relationships between the boolean presence of the batches, but also impose additional relationships between their start times, which are only enforced if the batches are actually present. This is achieved through reification. Thus, these SB constraints modify the nice boolean structure that makes the SAT methods in CP-SAT shine, exchanging it with reified constraints that negatively impact the models' performance. This behavior has been documented in the CP-SAT Primer by \citet{Krupke_The_CP-SAT_Primer}.

The best \rt{relative} gaps in CP-SAT belong to the AS model, followed by the RS model, very close to 0\%. Interestingly, the very same reason that makes The AS model model fall back in CP Optimizer, makes it shine in CP-SAT instead. The clear distinction between assignment and scheduling logic is well handled in CP-SAT thanks to the SAT component of this solver. Additionally, the RS model consistently produces comparable gaps of the AS model. However, as seen in Figure \ref{fig: CP-SAT TWCT no solution}, the RS model is not able to find solutions in all the instances.
}

\nt{
\subsection{Results with the makespan objective}

Figures \ref{fig: CP Opt Cmax} and \ref{fig: CP-SAT Cmax} present the results for minimizing $\makespan$ using CP Optimizer and CP-SAT, respectively. The results clearly indicate that the $\makespan$ objective is easier to solve than TWCT, primarily because it is insensitive to the processing order of jobs—only the maximum completion time matters. As a consequence, several models are able to prove optimality in some instances, as shown in Figures \ref{fig: CP Opt Cmax optimality} and \ref{fig: CP-SAT Cmax optimality}, and they achieve these optimal solutions before the time limit is reached, as evidenced by the solve times in Figures \ref{fig: CP Opt Cmax Time} and \ref{fig: CP-SAT Cmax Time}. Interestingly, the seasonal behavior observed in these figures suggests that instances with a higher number of machines are easier to solve, since fewer jobs per machine simplify the identification of the maximum completion time. In CP Optimizer, Figures \ref{fig: CP Opt Cmax optimality} and \ref{fig: CP Opt Cmax Time} further reveal that the A model scales exceptionally well, finding optimal solutions in a larger fraction of instances and in less time—thanks largely to the effective domain pruning provided by the $\AlwaysEqual$ global constraint with alignment parameters.

Figures \ref{fig: CP Opt Cmax no solution} and \ref{fig: CP-SAT Cmax no solution} (percentage of instances without a solution) and Figures \ref{fig: CP Opt Cmax Gap} and \ref{fig: CP-SAT Cmax Gap} (average gap with 95\% CI) tell a story similar to that of the TWCT objective. In CP Optimizer, the clear separation between assignment and scheduling decisions in the AS and AS+SB models hampers performance, whereas this separation benefits CP-SAT. In both solvers, the RS+SB model struggles, as the additional symmetry-breaking constraints become increasingly difficult to satisfy. The AS+SB model exhibits a similar issue: although the SB constraints help prove optimality for small instances in CP Optimizer, they become counterproductive as the instance sizes grow, leading to poorer gap values. In CP-SAT, the necessary reification in these constraints generally degrades performance.

The AU model performs well for instances with up to 150 jobs; however, when scaling to 200 jobs, it faces challenges in finding solutions for instances with 8 or 10 machines. Finally, the RS and S models show average performance: they do not match the AS model in CP-SAT, yet they consistently outperform the RS+SB model.

}
\section{Concluding remarks} \label{sec: conclusion}

This paper addressed the incompatible case of parallel batch scheduling, where compatible jobs belong to the same family, and jobs of different families cannot be processed together in the same batch. The \nt{state-of-the-art CP model for this problem—the Aligned (A) model—relies on state functions, global constraints, and alignment parameters available only in a commercial CP solver, making it solver-dependent. To overcome this limitation, we have presented four new CP models that can be implemented in other (open-source) solvers:
\begin{itemize}
    \item The Automaton (AU) model, which relies on automaton constraints, available in various open-source solvers.
    \item The Assign-and-Schedule (AS) model, which extends the model by \citet{Nascimento2023} to handle multiple incompatible job families and includes additional constraints to enforce batch size requirements. This model distinctly separates assignment decisions (handled by binary variables) from batch sequencing decisions (handled by CP scheduling logic).
    \item The Synchronized (S) model, which extends the model by \citet{Cakici2024} to accommodate incompatible families by replacing binary assignment variables with interval variables that are synchronized with batch sequencing decisions.
    \item The Redundant Synchronized (RS) model, which exclusively uses global constraints for both assignment and scheduling decisions and introduces redundancy into the problem structure.
\end{itemize}

Extensive computational experiments on 1,000 instances under two objectives (TWCT and $\makespan$) and across two solvers (CP Optimizer and CP-SAT) showcase the implementation flexibility of our proposed models. Our results reveal that:
\begin{itemize}
    \item The A model consistently produces high-quality solutions (0\% \rt{relative} gap) but is limited by its reliance on proprietary solver features only available in the commercial solver CP Optimizer.
    \item The clear separation between binary assignment and batch scheduling decisions enables the AS model to excel in the open-source solver CP-SAT. Thanks to the satisfiability components of the solver, which efficient\rt{ly} handle binary structures, the AS model achieves very small gaps and even proves optimality in isolated cases.
    \item The same separation that benefits the AS model in CP-SAT, however, undermines its performance in CP Optimizer, where global constraints lead to stronger propagation.
    \item The AU model shows promising scalability for moderate-sized instances, while the RS model, with its redundant structure, offers a competitive trade-off between propagation strength and computational efficiency.
    \item Incorporating symmetry-breaking (SB) constraints in the AS model under CP Optimizer is beneficial for proving optimality when time permits, but under tight time limits, these constraints become harder to satisfy and negatively impact solution quality.
    \item In CP-SAT, the SB constraints in the AS model, implemented via reification, introduce overhead in larger instances, thereby reducing the quality of the solutions obtained.
    \item The SB constraints in the RS model negatively impact its performance because they add additional  overhead that complicates the propagation process, leading to more frequent search failures and increased solution times in larger instances.
    \item Minimizing the makespan is considerably easier than minimizing the TWCT, mainly because the makespan objective is insensitive to the processing order and job weights. Instead, it focuses solely on the maximum completion time across all jobs, simplifying the scheduling decisions and often allowing models to prove optimality within the time limit.
\end{itemize}

By delivering additional solver-independent CP formulations, this paper not only extends the state-of-the-art for incompatible p-batch scheduling but also opens avenues for integrating these models into broader manufacturing scheduling contexts.} Current research includes using CP for serial batching, while future work will focus on extending efficient arc-flow formulations for compatible p-batch \cite{Trindade2018, Trindade2020, Trindade2021, Muter2020} to the incompatible case and embedding the proposed formulations within larger scheduling frameworks \cite{Fu2011, Ham2016}.

\section*{Acknowledgements} \label{sec: acknowledgments}
The authors would like to express their gratitude to Amira Hijazi for her support and helpful discussions. \nt{Additionally, to Petr Vilím, Eray Cakici, and the anonymous reviewers for their helpful comments that greatly improved the quality of the paper.}

% \ifCLASSOPTIONcaptionsoff
%   \newpage
% \fi
\bibliographystyle{IEEEtranN}

\bibliography{references}

\end{document}